\documentstyle[preprint,tighten,eqsecnum,prd,aps]{revtex}
\newcommand{\mysection}{\setcounter{equation}{0}\section}

\begin{document}
%for PACS nums -- of revtex
\draft

%-----------------------------
% ****** Start of file apssamp.tex ******
%
%   This file is part of the APS files in the REVTeX 3.1 distribution.
%   Version 3.1 of REVTeX, September 1996.
%
%   Copyright (c) 1992 The American Physical Society.
%
%   See the REVTeX 3.1 README file for restrictions and more information.
%
%
%

%ANTON: this  line works  only for preprint style
\preprint{INLO-PUB-22/99, YITP-SB-99-70}

\title{Bottom quark electroproduction in\\
variable flavor number schemes
}
\author{A. Chuvakin, J. Smith }
\address{
C.N. Yang Institute for Theoretical Physics,\\
State University of New York at Stony Brook,\\
New York 11794-3840, USA.\\
}
\author{W.L. van Neerven }
\address{
Instituut-Lorentz,\\
University of Leiden, PO Box 9506, 2300 RA Leiden,\
The Netherlands.
}
\date{\today}
\maketitle
\begin{abstract}
Two variable flavor number schemes are used to 
describe bottom quark production in deep inelastic electron-proton
scattering. In these schemes the coefficient functions are derived 
from mass factorization of the heavy quark coefficient functions 
presented in a fixed flavor number scheme. Also one has to
construct a parton density set with five light flavors (u,d,s,c,b)
out of a set which only contains four light flavors (u,d,s,c). In order
$\alpha_s^2$ the two sets are discontinuous at $\mu=m_b$ which follows from 
mass factorization of the heavy quark coefficient functions when it is
carried out in the ${\overline {\rm MS}}$-scheme.
Both variable flavor number schemes give almost identical
predictions for the bottom structure functions $F_{2,b}$ and $F_{L,b}$.
Also they both agree well with the corresponding results 
based on fixed order four-flavor perturbation theory 
over a wide range in $x$ and $Q^2$.
\end{abstract}
\pacs{PACS 11.10Jj, 12.38Bx, 13.60Hb, 13.87Ce.}

%anton
%\vfill

%\format=latex
%\documentstyle[12pt]{article}
%\pagestyle{myheadings}  
%\begin{document}
%------------------This is Section 1---------------------------------
\mysection{Introduction}
%----------------------------------------------------------
%\topmargin=0in
%\headheight=0in
%\headsep=0in
%\oddsidemargin=7.2pt
%\evensidemargin=7.2pt
%\footheight=1in
%\marginparwidth=0in
%\marginparsep=0in
%\textheight=9in
%\textwidth=6in
\newcommand{\be}{\begin{eqnarray}}
\newcommand{\ee}{\end{eqnarray}}

The H1 and ZEUS experiments at HERA now
have enough integrated luminosity to study bottom quark deep inelastic
electroproduction. 
Therefore it is an appropriate time to present up-to-date predictions for the
bottom quark components of the deep inelastic structure functions
$F_i(x,Q^2)$ where $i=2,L$.
We assume that the bottom quark is produced in an extrinsic fashion and
that the neutral current reaction dominates over the charged current 
one. This means that in fixed order perturbative QCD 
the heavy quark structure functions
$F_{i,b}(x,Q^2,m_b^2)$, $i=2,L$ are given by the 
virtual-photon gluon-fusion processes and their higher order
corrections with only light partons in the initial state. 
Notice that in the case of bottom quark production the light partons are 
represented by the gluon and the four light flavors u, d, s, c together 
with their anti-particles. 

In the literature one has adopted two different treatments of 
extrinsic bottom quark production, which are known as the massive and 
massless descriptions. The former treats the
bottom quark as a heavy quark (with mass $m_b$) 
and the partonic cross sections (or heavy quark coefficient functions)
are described by fixed order perturbation theory (FOPT)
as mentioned above. Notice that due to
the work in \cite{lrsn} the perturbation series is now known up to second
order. The latter treatment, which has been rather popular 
among groups which fit parton densities to experimental data, 
treats the bottom quark as a massless quark
so that it can be represented by a scale dependent 
parton density $f_b(x,\mu^2)$. Although at first
sight these approaches are completely different they are actually
intimately related. It was shown in \cite{bmsn1} that the
large logarithms of the type $\ln(Q^2/m_b^2)$, which appear in FOPT 
when $Q^2 \gg m_b^2$,
can be resummed in all orders. The upshot of this 
procedure is that the bottom components of the deep inelastic structure 
functions $F_{i,b}(x,Q^2,m_b^2)$, where $i=2,L$, which in the 
FOPT approach are written as convolutions of heavy quark coefficient 
functions with four-flavor light-mass (u,d,s,c) parton densities, become,
after resummation, convolutions of light-mass parton coefficient functions
with five-flavor light-mass parton densities 
which also include a bottom quark density. 
This procedure leads to the so-called zero mass variable flavor number scheme
(ZM-VFNS) for $F_{i,b}(x,Q^2)$ where the mass 
of the bottom quark is absorbed into the new five flavor densities. 
To implement this scheme one has to be careful to use quantities which
are collinearly finite in the limit $m_b \rightarrow 0$.
From the above considerations it is clear that the FOPT 
approach is better when the bottom quark pair is
produced near threshold (where $Q^2(x^{-1}-1)\ge 4 m_b^2$)
because terms in $m_b$ are important in this kinematic region.
On the other hand far above threshold, where also $Q^2\gg m_b^2$, 
the large logarithms mentioned above dominate the structure 
functions so that the ZM-VFNS approach should be more appropriate. Both
approaches are characterized by the number of active flavors involved in the
description of the parton densities which are given by four and five 
respectively. Each scheme has different 
densities but the momentum sum rule either gets contributions from
four-flavor densities or five-flavor densities and is always satisfied. 

As most of the experimental data will be in the kinematical regime which
is between the threshold and the large $Q^2$ region,  
a third approach, called the variable flavour number scheme (VFNS), 
has been introduced to describe the heavy quark components 
$F_{i,b}(x,Q^2)$ of the deep inelastic structure functions.
Actually there are several such schemes. They include the 
Aivasis, Collins, Olness, Tung (ACOT) \cite{acot},
scheme, the Buza, Matiounine, Smith, van Neerven 
(BMSN) \cite{bmsn1}, \cite{bmsn2} scheme, the Thorne, Roberts (TR) 
\cite{thro} scheme and the 
Chuvakin, Smith, van Neerven (CSN) \cite{csn1} scheme.
A discussion of them is given in the last reference.
The difference between the schemes can be attributed to
two ingredients entering in their construction.
The first one is the mass factorization procedure carried out before the 
large logarithms can be resummed. The second one is the matching condition 
imposed on the heavy quark density, which has to vanish in the threshold 
region of the production process. 
Another aspect of these approaches is that one needs two sets of 
parton densities. For bottom quark production one set only
contains densities in a four-flavor number scheme whereas the second one,
which also includes a bottom quark density, is parametrized in a five-flavor 
number scheme. Both parameterizations have to satisfy the 
matching relations quoted in \cite{bmsn1}. Up to next-to-leading order (NLO)
they are continuous at the scale $\mu=m_b$ whereas in next-to-next-to
leading order (NNLO) the parton densities become discontinuous while
going from a four to a five flavor scheme.
Starting from a three-flavor set of parton densities 
given in \cite{grv98} we have recently constructed in \cite{cs1}
a four-flavor set of densities which satisfied the matching relations in 
\cite{bmsn1} at the scale $\mu = m_c$. Then we evolved these
densities with LO or NLO splitting functions up to the scale $\mu = m_b$ and
constructed a five-flavor set which also satisfied the 
matching relations in \cite{bmsn1}. This set was further evolved
with LO and NLO splitting functions up to high scales. Notice that since
the NNLO splitting functions are unknown the only difference between the
NLO and NNLO parton densities can be attributed to the boundary conditions
at $\mu=m_c$ and $\mu=m_b$ where the latter densities become discontinuous 
contrary to the LO and NLO ones.
We can now use these densities to discuss VFNS 
for bottom quark deep inelastic electroproduction, in particular in the
CSN and BMSN schemes. 
The previous discussions in \cite{csn1}
were focussed on applications to charm quark electroproduction.
 
Since any description for bottom quarks follows
closely that for charm quarks we refer the interested reader to \cite{csn1}
for most of the details and simply specialize to bottom quark
electroproduction in Sec.II. We work to second order in the running
coupling constant $\alpha_s(\mu^2)$. Numerical results are shown for
the structure functions $F_{2,b}$ and $F_{L,b}$ in the CSN and BMSN schemes.

%\end{document}

%%%%%%%%%%%%%%%%%%%%%%%%%%%%%%%%%%%%%%%%%%%%%%%%%%%%%%%%%%%%%%%%%%%%%%%%
%\format=latex
%\documentstyle[12pt]{article}
%\pagestyle{myheadings}  
%\begin{document}
%------------------This is Section 2---------------------------------
\mysection{Bottom quark structure functions}
%----------------------------------------------------------
%\topmargin=0in
%\headheight=0in
%\headsep=0in
%\oddsidemargin=7.2pt
%\evensidemargin=7.2pt
%\footheight=1in
%\marginparwidth=0in
%\marginparsep=0in
%\textheight=9in
%\textwidth=6in 26 
%\newcommand{\be}{\begin{eqnarray}}
%\newcommand{\ee}{\end{eqnarray}}
\newcommand{\ssim}{\scriptstyle \sim}

In this Section we consider bottom quark deep inelastic electroproduction
in two variable flavor number schemes, namely the BMSN scheme as proposed in 
\cite{bmsn1}, \cite{bmsn2} and in the CSN scheme as proposed in
\cite{csn1}. For that purpose we have constructed in \cite{cs1} a
five-flavor parton density set from a four-flavor parton density set. 
Using our densities we
will study the differences between the bottom components of the 
deep inelastic structure functions $F_{i,b}^{\rm CSN}(n_f+1)$  
and $F_{i,b}^{\rm BMSN}(n_f+1)$, where the number of light flavors is
$n_f = 4$. 

To keep the discussion short we simply refer the reader to
Sec.III of \cite{csn1} for a discussion of the 
$\overline{\rm MS}$ parton densities,
the exact solution for the running coupling constant and the scale choice.
All references to three-flavour (four-flavor) densities should be
replaced by four-flavor (five-flavor) densities respectively.
All our calculations of next-to-leading (NLO) and next-to-next-leading
order (NNLO) quantities use 
$\Lambda_{3,4,5,6}^{\overline{\rm MS}}=299.4, 246, 167.7, 67.8~{\rm MeV}$,
which yields $\alpha_s(5,M_Z^2)=0.114$.
The structure functions are defined in Eqs.(3.9)-(3.17) of \cite{csn1}, where 
now $n_f = 4$ and $m_c$ is replaced by $m_b = 4.5$ ${\rm GeV/c}^2$.
To make this paper reasonably self-contained we
now reproduce the final formulae we use for the structure functions.
For $i=2,L$ the CSN scheme uses 
\begin{eqnarray}
\label{eqn2.1}
&& F_{i,Q}^{\rm CSN} (n_f+1,\Delta,Q^2,m^2)  =  e_Q^2 \left [
f_{Q +\bar Q}^{\rm NNLO}(n_f+1,\mu^2) {\cal C}_{i,Q}^{\rm CSN,NS,(0)}
(\frac{Q^2}{m^2}) \right.
\nonumber\\[2ex]
&& \left. +a_s(n_f+1,\mu^2) \left \{ f_{Q +\bar Q}^{\rm NLO}(n_f+1,\mu^2)
\otimes
{\cal C}_{i,Q}^{\rm CSN,NS,(1)}\Big (\frac{Q^2}{m^2},\frac{Q^2}{\mu^2}\Big )
\right. \right.
\nonumber\\[2ex]
&& \left. \left. +f_g^{\rm S,NLO}(n_f+1,\mu^2) \otimes
{\cal C}_{i,g}^{\rm CSN,S,(1)} \Big (\frac{Q^2}{m^2},\frac{Q^2}{\mu^2}
\Big ) \right \}
\right.
\nonumber\\[2ex]
&& \left. +a_s^2(n_f+1,\mu^2) \left \{
f_{Q +\bar Q}^{\rm LO}(n_f+1,\mu^2)\otimes \left (
{\cal C}_{i,q}^{\rm NS,(2)}
\Big (n_f+1,\frac{Q^2}{m^2}, \frac{Q^2}{\mu^2}\Big ) \right. \right. \right.
\nonumber\\[2ex]
&& \left. \left. \left. + {\cal C}_{i,q}^{\rm PS,(2)}
\Big(\frac{Q^2}{m^2}, \frac{Q^2}{\mu^2}\Big) \right )
 + \sum_{l=1}^{n_f}f_{l + \bar l}^{\rm LO} (n_f+1,\mu^2) \otimes
{\cal C}_{i,q}^{\rm CSN,PS,(2)}\Big(\frac{Q^2}{m^2}, \frac{Q^2}{\mu^2}\Big)
\right. \right.
\nonumber\\[2ex]
&& \left. \left. + f_g^{\rm S,LO}(n_f+1,\mu^2)\otimes
{\cal C}_{i,g}^{\rm CSN,S,(2)}
\Big (\frac{Q^2}{m^2},\frac{Q^2}{\mu^2} \Big )\right \}\right ]
\nonumber\\[2ex]
&& + a_s^2(n_f+1,\mu^2) \sum_{k=1}^{n_f}\,e_k^2 \,
f_{k + \bar k}^{\rm LO}(n_f+1,\mu^2)
\otimes L_{i,q}^{\rm HARD,NS,(2)} \Big (\Delta,\frac{Q^2}{m^2}\Big ) \,.
\end{eqnarray}
where we choose the heavy quark $Q$ to be the bottom quark,
so that the number of light flavors 
is $n_f=4$. The charge of the bottom quark is $e_Q=-1/3$ and its mass 
is $m = m_b$. 
The coefficient function $L_{i,q}^{\rm HARD}$ depends
on a parameter $\Delta$ which refers to the invariant mass of 
the $Q~\bar Q$-pair. For bottom quark production we choose 
$\Delta = 100$ $({\rm GeV}/c)^2$.
The coefficient functions labelled by $C^{\rm CSN}$ depend
on the heavy quark mass but are finite in the limit $m\rightarrow 0$.
They are defined in Eqs.(2.8)-(2.20) in \cite{csn1}.
To simplify the notation we will refer to the
above structure functions as $F_{i,b}^{\rm CSN}(n_f=5)$, indicating
that they depend on five-flavor parton densities. 

The same parameters $e_Q$, $m_b$, $\Delta$ etc., 
also show up in the expressions for the bottom quark structure
functions in the BMSN scheme. Here we have the representations
\begin{eqnarray}
\label{eqn2.2}
&&F_{i,Q}^{\rm BMSN}(n_f+1,\Delta,Q^2,m^2)=
F_{i,Q}^{\rm EXACT}(n_f,\Delta,Q^2,m^2)
\nonumber\\[2ex]
&& - F_{i,Q}^{\rm ASYMP}(n_f,\Delta,Q^2,m^2)
+ F_{i,Q}^{\rm PDF}(n_f+1,\Delta,Q^2,m^2)\,.
\end{eqnarray}
The pieces in this formulae represent first the results in FOPT, given by
\begin{eqnarray}
\label{eqn2.3}
&&F_{i,Q}^{\rm EXACT}(n_f,\Delta,Q^2,m^2) =
 e_Q^2 \left [ a_s(n_f,\mu^2) f_g^{\rm S,NLO}(n_f,\mu^2) \otimes
H_{i,g}^{\rm S,(1)}\Big(\frac{Q^2}{m^2}\Big) \right.
\nonumber\\[2ex]
&&\left. + a_s^2(n_f,\mu^2)\left \{ \sum_{k=1}^{n_f} f_{k + \bar k}^{\rm LO}
(n_f,\mu^2) \otimes
H_{i,q}^{\rm PS,(2)}\Big(\frac{Q^2}{m^2},\frac{Q^2}{\mu^2}\Big) \right. \right.
\nonumber\\[2ex]
&& \left. \left. +  f_g^{\rm S,LO}(n_f,\mu^2) \otimes
H_{i,g}^{\rm S,(2)}\Big(\frac{Q^2}{m^2},\frac{Q^2}{\mu^2}\Big) \right \}
\right ]
\nonumber\\[2ex]
&& + a_s^2(n_f,\mu^2)\sum_{k=1}^{n_f} e_k^2 f_{k+\bar k}^{\rm LO}(n_f,\mu^2)
\otimes L_{i,q}^{\rm HARD,NS,(2)} \Big(\Delta,\frac{Q^2}{m^2}\Big )\,.
\end{eqnarray}
The next pieces are the structure functions $F_{i,Q}^{\rm ASYMP}(n_f)$
which can be obtained from $F_{i,Q}^{\rm EXACT}(n_f)$ by
replacing all exact heavy quark coefficient functions $H_{i,k}$
and $ L_{i,k}$ ($k=q,g$) by their asymptotic analogues which are defined
by
\begin{eqnarray}
\label{eqn2.4}
H_{i,k}^{\rm ASYMP} = \mathop{\mbox{lim}}\limits_{\vphantom{
\frac{A}{A}} Q^2 \gg m^2} H_{i,k} \quad , \quad
L_{i,k}^{\rm ASYMP}= \mathop{\mbox{lim}}\limits_{\vphantom{
\frac{A}{A}} Q^2 \gg m^2} L_{i,k}\,. 
\end{eqnarray}
Finally the structure functions $F_{i,Q}^{\rm PDF}(n_f+1)$ which are 
very often called the ZM-VFNS representations are defined by
\begin{eqnarray}
\label{eqn2.5}
&& F_{i,Q}^{\rm PDF} (n_f+1,\Delta,Q^2,m^2)  =  e_Q^2 \Biggl [
f_{Q +\bar Q}^{\rm NNLO}(n_f+1,\mu^2) {\cal C}_{i,q}^{\rm NS,(0)}
\nonumber\\[2ex]
&& +a_s(n_f+1,\mu^2) \left \{ f_{Q +\bar Q}^{\rm NLO}(n_f+1,\mu^2)
\otimes {\cal C}_{i,q}^{\rm NS,(1)}(\frac{Q^2}{\mu^2}) \right.
\nonumber\\[2ex]
&& \left. +f_g^{\rm S,NLO}(n_f+1,\mu^2)\otimes \tilde
{\cal C}_{i,g}^{\rm S,(1)} (\frac{Q^2}{\mu^2}) \right \}
\nonumber\\[2ex]
&& +a_s^2(n_f+1,\mu^2) \left \{
f_{Q +\bar Q}^{\rm LO}(n_f+1,\mu^2)\otimes \left ({\cal C}_{i,q}^{\rm NS,(2)}
(n_f+1,\frac{Q^2}{\mu^2}) \right. \right.
\nonumber\\[2ex]
&& \left. \left.
+ \tilde {\cal C}_{i,q}^{\rm PS,(2)}\Big(\frac{Q^2}{\mu^2}\Big)\right)
+ \sum_{l=1}^{n_f}f_{l + \bar l}^{\rm LO} (n_f+1,\mu^2) \Big )\otimes
\tilde {\cal C}_{i,q}^{\rm PS,(2)}\Big(\frac{Q^2}{\mu^2}\Big)
\right.
\nonumber\\[2ex]
&&\left. + f_g^{\rm S,LO}(n_f+1,\mu^2)\otimes \tilde
{\cal C}_{i,g}^{\rm S,(2)} (\frac{Q^2}{\mu^2})\right \}\Biggr ]
\nonumber\\[2ex]
&& + a_s^2(n_f+1,\mu^2)\sum_{k=1}^{n_f} e_k^2 f_{k+\bar k}^{\rm LO}(n_f,\mu^2)
\otimes L_{i,q}^{\rm HARD,ASYMP,NS,(2)} \Big(\Delta,\frac{Q^2}{m^2}\Big )\,.
\nonumber\\[2ex]
\end{eqnarray}
In all these results the heavy quark Q refers to the bottom quark and the other
parameters are defined above.
For simplicity we will refer to these structure functions as 
$F_{i,b}^{\rm EXACT}(n_f = 4)$, $F_{i,b}^{\rm ASYMP}(n_f = 4)$,
and $F_{i,b}^{\rm PDF}(n_f = 5)$ respectively, which indicates that the
first two structure functions depend on four-flavor densities and the last one 
depends on five-flavor densities.
The parton densities $f_k$ in the above formulae are represented in leading 
order (LO), next-to-leading order (NLO) and next-to-next-to-leading order
(NNLO). The NNLO case refers
to the boundary conditions imposed in \cite{csn1} since the three-loop
splitting functions are not known yet. These parton densities have 
been constructed in \cite{cs1} starting from a three-flavor parametrization
in \cite{grv98}. The multiplication of the
densities with the heavy and light parton coefficient functions is 
done in such a way that the perturbation series is strictly truncated
at order $\alpha_s^2$. This is necessary to avoid scheme dependent terms which
would otherwise arise beyond order $\alpha_s^2$. Therefore the following
requirement is satisfied 
\begin{eqnarray}
\label{eqn2.6}
F_{i,Q}^{\rm CSN}(n_f=5) = F_{i,Q}^{\rm BMSN}(n_f=5) 
= F_{i,Q}^{\rm EXACT}(n_f=4) \quad \mbox{for} 
\quad Q^2 \le m^2 \,.
\end{eqnarray}
Since $f_Q(m^2)^{\rm NNLO} \not = 0$ (see \cite{bmsn1}) this condition can be 
only satisfied when we truncate the perturbation series at the same order.
Furthermore because of Eq. (\ref{eqn2.4}) and the property
\begin{eqnarray}
\label{eqn2.7}
\mathop{\mbox{lim}}\limits_{\vphantom{ \frac{A}{A}} Q^2 \gg m^2}
{\cal C}_{i,k}^{\rm CSN,(l)}(n_f+1,\frac{Q^2}{m^2},\frac{Q^2}{\mu^2})
={\cal C}_{i,k}^{(l)}(n_f+1,\frac{Q^2}{\mu^2}) \,,
\end{eqnarray}
we have the asymptotic relation
\begin{eqnarray}
\label{eqn2.8}
&& \mathop{\mbox{lim}}\limits_{\vphantom{ \frac{A}{A}} Q^2 \gg m^2}
F_{i,Q}^{\rm BMSN}(n_f+1,\Delta,Q^2,m^2)
= \mathop{\mbox{lim}}\limits_{\vphantom{
\frac{A}{A}} Q^2 \gg m^2} F_{i,Q}^{\rm CSN}(n_f+1,\Delta,Q^2,m^2)
\nonumber\\[2ex]
&& = F_{i,Q}^{\rm PDF}(n_f+1,\Delta,Q^2,m^2) \,.
\end{eqnarray}
At first sight the form of the expression for $F_{i,Q}^{\rm BMSN}$ in
Eq. (\ref{eqn2.2}) looks quite different from the one presented
for $F_{i,Q}^{\rm CSN}$ in Eq. (\ref{eqn2.1}). However this is not true.
Using the mass factorization relations for the asymptotic heavy
quark coefficient functions in \cite{bmsn1} one can cast 
$F_{i,Q}^{\rm BMSN}$ into the same form as presented for
$F_{i,Q}^{\rm CSN}$ where all quark coefficient functions of the type
${\cal C}_{i,Q}^{\rm CSN}$ are replaced by their light quark analogues
${\cal C}_{i,q}$ appearing in Eq. (\ref{eqn2.5}). This replacement
also applies to the ${\cal C}_{i,Q}^{\rm CSN}$ occurring in the mass
factorization relations for ${\cal C}_{i,g}^{\rm CSN,S}$ and
${\cal C}_{i,q}^{\rm CSN,PS}$ presented in \cite{csn1}. 
Therefore the difference between the CSN and BMSN schemes can be attributed
to the powers $(m^2/Q^2)^j$ showing up in ${\cal C}_{i,Q}^{\rm CSN}$
but absent in ${\cal C}_{i,q}$. This effect is only noticeable
in the threshold region where $Q^2 \sim m^2$ as we will show below.

The heavy quark coefficient functions  ${\cal C}_{i,k}^{\rm CSN}$, 
$H_{i,k}$, $L_{i,k}$ ($k=Q,q,g$) and the
light partonic coefficient functions ${\cal C}_{i,k}$ ($k=q,g$)
can be traced back to the following processes
\begin{eqnarray}
\label{eqn2.9}
{\cal C}_{i,g}^{\rm CSN,S,(1)}, H_{i,g}^{\rm S,(1)} 
&:& \gamma^* + g \rightarrow Q + \bar Q
\quad \mbox{\cite{csn1} (CSN), \cite{lrsn}~~(EXACT),}
\nonumber\\[1ex]
&&\hspace{40mm} \mbox{\cite{bmsmn}~~(ASYMP)}
\nonumber\\[2ex]
{\cal C}_{i,g}^{\rm CSN,S,(2)}, H_{i,g}^{\rm S,(2)} 
&:& \gamma^* + g \rightarrow Q + \bar Q + g
\quad \mbox{\cite{csn1} (CSN), \cite{lrsn}~~(EXACT),}
\nonumber\\[1ex]
&&\hspace{40mm} \mbox{\cite{bmsmn}~~(ASYMP)}
\nonumber\\[2ex]
{\cal C}_{i,q}^{\rm CSN,PS,(2)}, H_{i,q}^{\rm PS,(2)} 
&:& \gamma^* + q \rightarrow Q
+ \bar Q + q
\quad \mbox{Bethe-Heitler reaction}
\nonumber\\[1ex]
&&  \mbox{\cite{csn1}~~(CSN), \cite{lrsn}~~(EXACT), 
\cite{bmsmn}~~(ASYMP)}
\nonumber\\[2ex]
L_{i,q}^{\rm HARD,NS,(2)} &:& \gamma^*
+ q \rightarrow Q + \bar Q + q
\quad \mbox{Compton reaction}
\nonumber\\[1ex]
&& \hspace{35mm} \mbox{\cite{csn1} (EXACT and ASYMP)}
\nonumber\\[2ex]
{\cal C}_{i,Q}^{\rm CSN,NS,(0)}, H_{i,Q}^{\rm NS,(0)} 
&:& \gamma^* + Q \rightarrow Q
\nonumber\\[2ex]
{\cal C}_{i,Q}^{\rm CSN,NS,(1)}, H_{i,Q}^{\rm NS,(1)} 
&:&\gamma^* + Q \rightarrow Q + g
\quad \mbox{\cite{neve}} 
\nonumber\\[2ex]
{\cal C}_{i,q}^{\rm NS,(0)} &:&\gamma^* + q \rightarrow q 
\nonumber\\[2ex]
{\cal C}_{i,q}^{\rm NS,(1)} &:&\gamma^* + q \rightarrow q + g
\quad \mbox{\cite{zn}}
\nonumber\\[2ex]
{\cal C}_{i,q}^{\rm NS,(2)} &:&\gamma^* + q \rightarrow q + g + g
\quad \mbox{\cite{zn}}
\nonumber\\[2ex]
{\cal C}_{i,q}^{\rm NS,(2)} &:&\gamma^* + q \rightarrow q + \bar q + q
\quad \mbox{\cite{zn}}
\nonumber\\[2ex]
\tilde {\cal C}_{i,q}^{\rm PS,(2)} &:&\gamma^* + q \rightarrow q + \bar q + q
\quad \mbox{\cite{zn}}
\nonumber\\[2ex]
\tilde {\cal C}_{i,g}^{\rm S,(1)} &:&\gamma^* + g \rightarrow q + \bar q
\quad \mbox{\cite{zn}}
\nonumber\\[2ex]
\tilde {\cal C}_{i,g}^{\rm S,(2)} &:&\gamma^* + g \rightarrow q + \bar q + g
\quad \mbox{\cite{zn}}.
\end{eqnarray}
Behind the reactions we have quoted the references in which the
corresponding coefficient functions can be found. Note that the heavy quark
coefficient functions $H_{i,k}$ are mass singular when $m \rightarrow 0$. This
can be observed immediately when one looks at $H_{i,k}^{\rm ASYMP}$
which behaves like $\ln^m (\mu^2/m^2)~\ln^n (Q^2/m^2)$ (see \cite{bmsmn}).
After the logarithms are removed one obtains the quantities
${\cal C}_{i,k}^{\rm CNS}$ which, even though they depend
on $m$,  are finite in the limit $m \rightarrow 0$.
The coefficient function $L_{i,k}^{\rm HARD}$ is finite by itself because
as we mentioned above we have imposed a lower cut
off $\Delta = 100$ $({\rm GeV}/c)^2$
on the invariant mass of the $Q~\bar Q$-pair. 
Finally notice that all parton densities,
coefficient functions and the running coupling constant are presented
in the ${\overline {\rm MS}}$-scheme.

Now we present results for the various structure functions.
We are interested in the bottom quark structure functions 
$F_{i,b}^{\rm CSN}(n_f = 5)$
and $F_{i,b}^{\rm BMSN}(n_f=5)$ for $i=2,L$ in 
NNLO for the CSN \cite{csn1} and BMSN \cite{bmsn1} schemes
respectively.
In Fig. 1 we have plotted the structure functions $F_{2,b}^{\rm CSN}(n_f=5)$,
$F_{2,b}^{\rm BMSN}(n_f=5)$, $F_{2,b}^{\rm PDF}(n_f=5)$ and 
$F_{2,b}^{\rm EXACT}(n_f=4)$
in the region $20 < Q^2 < 10^3$ in units of $({\rm GeV/c})^2$
for $x=0.05$. This figure reveals that there is hardly
any difference between the BMSN and CSN prescriptions. The curves in both
prescriptions are essentially identical to  
that for $F_{2,b}^{\rm EXACT}(n_f=4)$.
In this region $F_{2,b}^{\rm PDF}(n_f=5)$
is larger than the other results which is expected from the 
discussion of the bottom quark density given in \cite{cs1}.
There is still an appreciable difference at the highest plotted $Q^2$ 
demonstrating that mass effects are important up to very large scales.
Notice that for $Q^2 \le 35$ $({\rm GeV/c})^2$ 
$F_{2,b}^{\rm PDF}(n_f=5)$ becomes negative which means that bottom
quark electroproduction cannot be described by this quantity anymore.
In Fig. 2 we present the same plots for $x=0.005$. Again one cannot
distinguish between $F_{2,b}^{\rm BMSN}(n_f=5)$ and 
$F_{2,c}^{\rm CSN}(n_f=5)$
but now both are smaller than $F_{2,b}^{\rm EXACT}(n_f=4)$ over the whole $Q^2$
range. The latter is smaller than $F_{2,b}^{\rm PDF}(n_f=5)$ in particular
for $Q^2>35~({\rm GeV/c})^2$.
Further we want to emphasize that due to our careful treatment of the
threshold region there is an excellent cancellation 
between $F_{2,b}^{\rm PDF}(n_f=5)$ and $F_{2,b}^{\rm ASYMP}(n_f=4)$ 
so that both $F_{2,b}^{\rm CSN}(n_f=5)$ 
and $F_{2,b}^{\rm BMSN}(n_f=5)$ 
tend to $F_{2,b}^{\rm EXACT}(n_f=4)$ at $Q^2 = m_b^2$. 
At large $Q^2$ we have a cancellation between 
$F_{2,b}^{\rm ASYMP}(n_f=4)$ and $F_{2,b}^{\rm EXACT}(n_f=4)$  
so that both $F_{2,b}^{\rm CSN}(n_f=5)$ 
and $F_{2,b}^{\rm BMSN}(n_f=5)$ slowly tend to 
$F_{2,b}^{\rm PDF}(n_f=5)$. They are only identical at extremely large
$Q^2$ demonstrating that mass effects are still important 
over a wide range in $x$ and $Q^2$.

In Fig.3 we show similar plots as in Fig.1 for the 
bottom quark longitudinal structure functions. 
Here we observe a small difference between the plots for
$F_{L,b}^{\rm CSN}(n_f=5)$ and $F_{L,b}^{\rm BMSN}(n_f=5)$ in the
region $20 < Q^2 < 10^3~({\rm GeV/c})^2$.  Furthermore
$F_{L,b}^{\rm PDF}(n_f=5)$ is considerably larger than the other three
structure functions, which differs from the behavior seen 
in Fig.1. This can be mainly attributed to the gluon density which 
plays a more prominant role in $F_{L,b}$ than in $F_{2,b}$. For $x=0.005$ 
(see Fig.4) the small difference between the BMSN and 
the CSN descriptions becomes more conspicuous for low $Q^2$. 

In Figs.5 and 6 we make a comparison between the NLO and the NNLO
structure functions $F_{2,b}^{\rm CSN}(n_f=5)$ and 
$F_{2,b}^{\rm BMSN}(n_f=5)$. Both the CSN and and BMSN descriptions lead to
the same results in both NLO and NNLO. However while going from NLO to 
NNLO the the structure functions $F_{2,b}^{\rm CSN}(n_f=5)$ and
$F_{2,b}^{\rm BMSN}(n_f=5)$ increase a little bit. The differences 
in the case of $x=0.005$ in Fig.6 are even smaller 
than those observed for $x=0.05$ in Fig.5.
The same comparison between NLO and NNLO results is made for the longitudinal
structure functions in Figs.7 and 8. Here the differences between
NLO and NNLO cases are much larger than in the case of $F_{2,b}$ in Figs.5,6.
In NLO both $F_{L,b}^{\rm BMSN}(n_f=5)$ and $F_{L,b}^{\rm CSN}(n_f=5)$ 
are smaller than the NNLO results. 

Previous results for $F_{i,b}^{\rm EXACT}(x,Q^2,m_b^2)$ and have been presented 
in Figs. 20a,20b in \cite{lrsn} for a now obsolete set of parton densities, 
so the values quoted there are too small. To show these changes  
we add in Figs.9 and 10 plots for the $x$ dependence of
$F_{2,b}^{\rm CSN}(n_f=5)$, $F_{2,b}^{\rm BMSN}(n_f=5)$, 
$F_{2,b}^{\rm PDF}(n_f=5)$ and $F_{2,b}^{\rm EXACT}(n_f=4)$  
at fixed $Q^2 = 30$ and $Q^2 = 100$ in units of $({\rm GeV/c})^2$
respectively. 
Finally we also show in Figs.11 and 12 plots for the $x$ dependence of
$F_{L,b}^{\rm CSN}(n_f=5)$, $F_{L,b}^{\rm BMSN}(n_f=5)$, 
$F_{L,b}^{\rm PDF}(n_f=5)$ and $F_{L,b}^{\rm EXACT}(n_f=4)$  
at fixed $Q^2 = 30$ and $Q^2 = 100$ in units of $({\rm GeV/c})^2$
respectively. 
Note that there are also some recent results
for $F_{2,b}$ in \cite{thro} in the TR scheme and in \cite{kd} 
for FOPT.

The plots for $F_{L,b}^{\rm CSN}(n_f=5)$ in Figs.7,8 do not show a 
negative region at small $Q^2$ which could have been expected by analogy 
with the results for $F_{L,c}^{\rm CSN}(n_f=4)$ in \cite{csn1}.
In the case of bottom quark production the negative regions do occur but
at even smaller values of $x$.
In Figs. 13, 14 we show the same plots as in Figs.1,3 respectively
but for $x=5\times 10^{-5}$.
Now the structure function $F_{L,b}^{\rm CSN}$
is negative in the region $Q^2 \approx 30$ $({\rm GeV/c})^2$.
In Figs. 15,16 we show the same plots as in Figs. 5,7 respectively
but for $x=5\times 10^{-5}$.  Fig. 16 shows that the longitudinal structure
functions for the case of bottom production
also have negative regions at small $Q^2$ values in
both NLO and NNLO. This phenomenon also occurs for the charm quark 
structure functions in \cite{csn1}.
In the NLO case this arises because  
the term $f_g^{\rm S,NLO}(n_f+1,\mu^2) \otimes {\cal C}_{L,g}^{\rm CSN,S,(1)}
\Big ({Q^2}/{m^2},{Q^2}/{\mu^2}\Big )$ in Eq. (\ref{eqn2.1})
is negative due to the definition of the gluon coefficient function
in the CSN scheme (see Eq.(2.19)) in \cite{csn1}) which is given by
\begin{eqnarray}
\label{eqn2.10}
{\cal C}_{L,g}^{\rm CSN,S,(1)}\Big (\frac {Q^2}{m^2},\frac {Q^2}{\mu^2}\Big)&=&
H_{L,g}^{\rm S,(1)}\Big (\frac {Q^2}{m^2}\Big)-
A_{Qg}^{\rm S,(1)}(\frac{\mu^2}{m^2})\, 
{\cal C}_{L,Q}^{\rm CSN,NS,(0)}\Big (\frac {Q^2}{m^2}\Big) \,,
\nonumber\\[2ex]
\mbox{with} \quad {\cal C}_{L,Q}^{\rm CSN,NS,(0)}&=&\frac{4m^2}{Q^2} \,,
\end{eqnarray}
where $A_{Qg}^{\rm S,(1)}$ denotes the one-loop operator matrix element
computed in \cite{bmsmn}. Notice that the latter and the lowest order 
exact coefficient function $H_{L,g}^{\rm S,(1)}$ are always positive. 
Because of the minus sign in Eq. (\ref{eqn2.10}) it appears that the 
coefficient function ${\cal C}_{L,g}^{\rm CSN,S,(1)}$ can become negative 
in particular at low $Q^2$ values. In the NNLO case one obtains more
negative contributions due to the term
$f_{Q +\bar Q}^{\rm NNLO}(n_f+1,\mu^2) {\cal C}_{L,Q}^{\rm CSN,NS,(0)}$ 
in formula (\ref{eqn2.1}). It turns out
that $f_{Q +\bar Q}^{\rm NNLO}(n_f+1,x,\mu^2)$ is negative at small $x$
and $\mu^2=Q^2 \ge m^2$. Notice that at the latter scale
$f_{Q +\bar Q}^{\rm LO}(n_f+1,x,\mu^2)$ and 
$f_{Q+\bar Q}^{\rm NLO}(n_f+1,x,\mu^2)$
are very small because they vanish at $\mu = m$ in contrast to
$f_{Q +\bar Q}^{\rm NNLO}(n_f+1,x,\mu^2)$. The behavior of the
structure function above is characteristic of the CSN scheme since
it does not appear in the case of BMSN.
This is because in the latter scheme the longitudinal
coefficient function, represented by ${\cal C}_{L,q}^{\rm CSN,NS,(0)}$,
is identical to zero so that the zeroth order contribution to
$F_{L,b}^{\rm BMSN}(n_f=5)$ vanishes and the first order correction
is given by ${\cal C}_{L,g}^{\rm BMSN,S,(1)}=H_{L,g}^{\rm S,(1)}$. 
The latter leads to a positive structure function over the whole kinematical
region.
To further demonstrate this point we plot in Fig. 17 pieces of the 
NLO result
\begin{eqnarray}
\label{eqn2.11}
&& F_{L,b}^{\rm CSN} (n_f+1,Q^2,m^2)  =  e_b^2 \left [
f_{b +\bar b}^{\rm NLO}(n_f+1,\mu^2) {\cal C}_{L,b}^{\rm CSN,NS,(0)}
(\frac{Q^2}{m^2}) \right.
\nonumber\\[2ex]
&& \left. +a_s(n_f+1,\mu^2) \left \{ f_{b +\bar b}^{\rm LO}(n_f+1,\mu^2)
\otimes
{\cal C}_{L,b}^{\rm CSN,NS,(1)}\Big (\frac{Q^2}{m^2},\frac{Q^2}{\mu^2}\Big )
\right. \right.
\nonumber\\[2ex]
&& \left. \left. +f_g^{\rm S,LO}(n_f+1,\mu^2) \otimes
{\cal C}_{L,g}^{\rm CSN,S,(1)} \Big (\frac{Q^2}{m^2},\frac{Q^2}{\mu^2}
\Big ) \right \}
\right. \,.
\end{eqnarray}
The sum of the b-quark contributions, labelled Term1, is always positive. 
The gluonic contribution, labelled Term2, is clearly negative over a wide
range in $Q^2$ and large enough that the sum $F_{L,b}^{\rm CSN}$
is also negative for $30 \le Q^2 \le 150$ ${\rm GeV}^2$, as in Fig.16. 
This behavior is due to the gluonic coefficient function 
${\cal C}_{L,g}^{\rm CSN,S,(1)}$ which is explained below Eq. (\ref{eqn2.10}).
However the magnitude of the gluonic contribution depends on the choice of the
gluon density. If we use an NLO gluon density in the order $\alpha_s$
contribution to the structure function in Eq. (\ref{eqn2.11}) rather than a LO
gluon density then the sum of the first two terms is unchanged
but the gluonic part is now smaller in magnitude. These contributions are shown
in Fig.18 where now $|{\rm Term2}| \le {\rm Term1}$ so that the total result 
for $F_{L,b}^{\rm CSN}$ is everywhere positive. However this procedure
violates our prescription for the computation of the structure functions
in both the CSN and the BMSN schemes. In this prescription the LO densities
are multiplied by the highest order coefficient function whereas the NLO
densities are combined with lower order coefficient functions (see formulae
(\ref{eqn2.1}), (\ref{eqn2.3}) and (\ref{eqn2.5})). In this way the
perturbation series is truncated up to the order we want to compute the
structure functions. Hence we avoid terms, arising beyond that order,
which introduce a scheme dependence and spoils the threshold behavior
(see \cite{csn1}). The latter happens if one follows the usual procedure
where one multiplies the highest order densities by the highest order
coefficient functions.
The difference between the usual procedure and our prescription is not
only shown by our parton density set but is also observed for other sets
presented in the literature. Examples are recent sets such as MRST98 
\cite{mrst98} (with $m_b = 4.3$ ${\rm GeV}$, $m_c = 1.35$ ${\rm GeV}$), 
MRST99 \cite{mrst99} (with $m_b = 4.3$ ${\rm GeV}$, $m_c = 1.43$ ${\rm GeV}$), 
and CTEQ5 \cite{cteq5}
(with $m_b = 4.5$ ${\rm GeV}$, $m_c = 1.3$ ${\rm GeV}$). 
Note that the MRST99 set does not provide LO densities. 
Using their NLO densities they yield positive values for the $Q^2$ 
dependence of $F_{L,b}^{\rm CSN}$ at $x= 5\times 10^{-5}$.
There are both LO (CTEQ5L) and NLO (CTEQ5M) densities in the
CTEQ5 set and we have checked that, 
for the same $x$, $Q^2$ values $F_{L,b}^{\rm CSN}$ is
positive with purely NLO densities but has a negative region when the
LO and NLO densities are used according to our prescription. 
The observations made above leads to the conclusion that the
$4 m^2/Q^2$ term in the non-singlet CSN
longitudinal coefficient function in Eq. (\ref{eqn2.10}) leads to a negative
gluonic coefficient function. When the latter is used
together with the latest LO and NLO parton density sets  
$F_{L,b}^{\rm CSN}(n_f=5)$ becomes negative in the low $Q^2$-region at small 
$x$. We speculate that the CSN scheme would always yield positive
structure functions if we could use 
parton density sets which fitted data with convolutions
of LO densities with $O(\alpha_s)$ coefficient functions and NLO densities 
with zeroth order coefficient functions. 
Unfortunately such densities are not available.

To summarize the main points we have implemented 
two variable flavor number schemes for bottom quark electroproduction
in NNLO and compared them with NLO FOPT results. 
The schemes differ in the way mass factorization
is implemented. In the CSN scheme this is done with respect to the
full heavy and light quark structure functions at finite $Q^2$. 
In the BMSN scheme the mass factorization is only applied to the
coefficient functions in the large $Q^2$ limit. Both schemes require
four-flavor and five-flavor parton densities which satisfy discontinuous NNLO
matching conditions at a scale $\mu = m_b$. We have constructed these 
densities using our own evolution code \cite{cs1}.
The schemes also require matching conditions on the coefficient functions 
which are implemented in this paper. Note that we have also removed the 
dangerous terms in $\ln^3(Q^2/m_b^2)$ from the Compton 
contributions so that both
$F_{i,b}^{\rm CSN}(n_f=5)$ and
$F_{i,b}^{\rm BMSN}(n_f=5)$ are collinear safe. As in \cite{csn1}
we have done this in a way which is consistent with our study of
inclusive quantities by implementing a cut $\Delta$ on the mass of 
the $b-\bar b$ pair. We stress that 
any ZM-VFNS bottom quark density description of $F_{i,b}$
must use collinear safe definitions. This is not required in the
fixed order perturbation theory approach given by $F_{i,b}^{\rm EXACT}(n_f=4)$
in \cite{lrsn} for moderate $Q^2$-values.

Finally we made a careful analysis of the threshold behaviors of
$F_{i,b}^{\rm CSN}(n_f=5)$ and
$F_{i,b}^{\rm BMSN}(n_f=5)$. In order to
achieve the required cancellations at the scale $\mu = m_b$
so that they both become
equal to $F_{i,b}^{\rm EXACT}(n_f=4)$ one must be 
very careful to combine terms with the same order in the
expansion in $\alpha_s$. The approximation 
we made in this paper, of using NLO splitting functions in place of
NNLO splitting functions, was sufficient for our purposes.
We successfully implemented the required cancellations near threshold and the
corresponding limits at large scales came out naturally. Inconsistent sets
of parton densities automatically spoil these cancelations. 
Since there are only minor differences between
the CSN, BMSN and NLO FOPT predictions
it is clear that the use of variable flavor number schemes for 
bottom quark production is not required for the analysis of HERA data.
However the ZM-VFNS description is clearly inadequate at small $Q^2$.

ACKNOWLEDGMENTS

The work of A. Chuvakin and J. Smith was partially supported
by the National Science Foundation grant PHY-9722101.
The work of W.L. van Neerven was supported
by the EC network `QCD and Particle Structure' under contract
No.~FMRX--CT98--0194.

%\end{document}

%\documentstyle[12pt]{article}
%\pagestyle{myheadings}  
%\begin{document}
%----------------------------References-------------------------------------
%

%\documentstyle[12pt]{article}
%\begin{document}
\centerline{\bf \large{Figure Captions}}
\begin{description}
%----------------------------------------
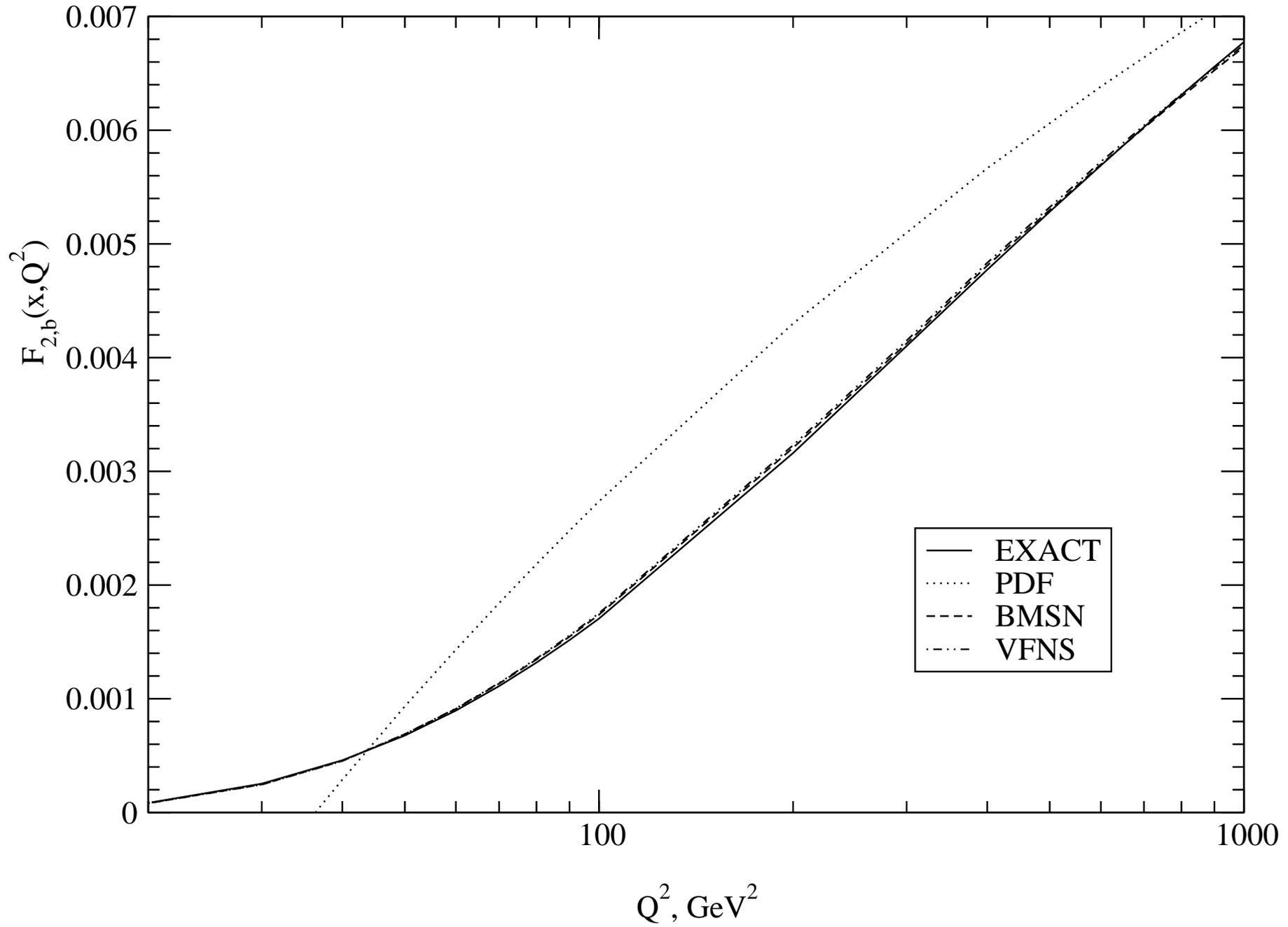
\item[Fig. 1.]
The bottom quark structure functions 
$F_{2,b}^{\rm EXACT}(n_f=4)$ (solid line) 
$F_{2,b}^{\rm CSN}(n_f=5)$, (dot-dashed line)  
$F_{2,b}^{\rm BMSN}(n_f=5)$, (dashed line) and
$F_{2,b}^{\rm PDF}(n_f=5)$, (dotted line) 
in NNLO for $x=0.05$ 
plotted as functions of $Q^2$.
%-------------------------------------------------
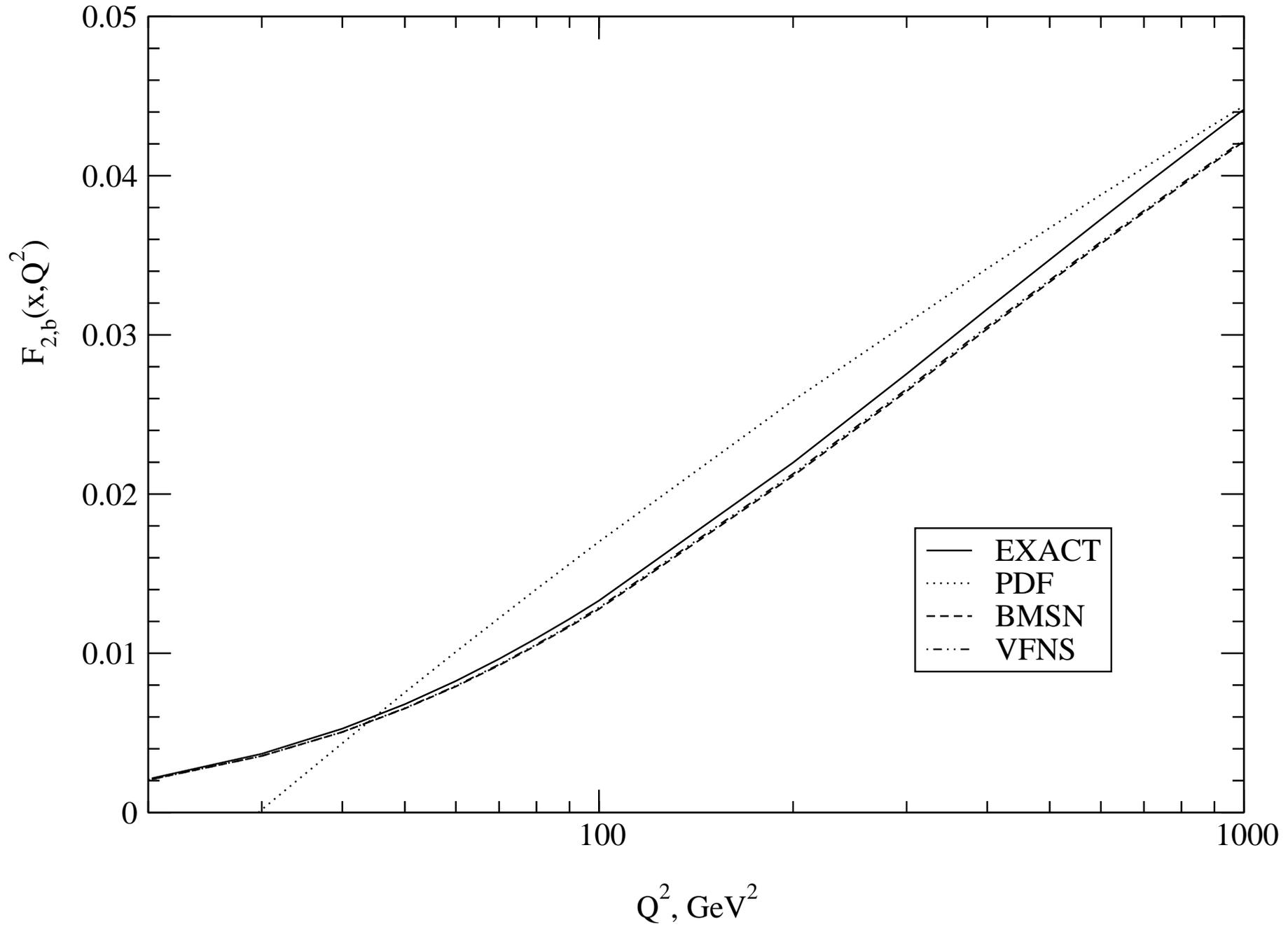
\item[Fig. 2.] Same as in Fig. 1 but now for $x=0.005$.
%-------------------------------------------------------
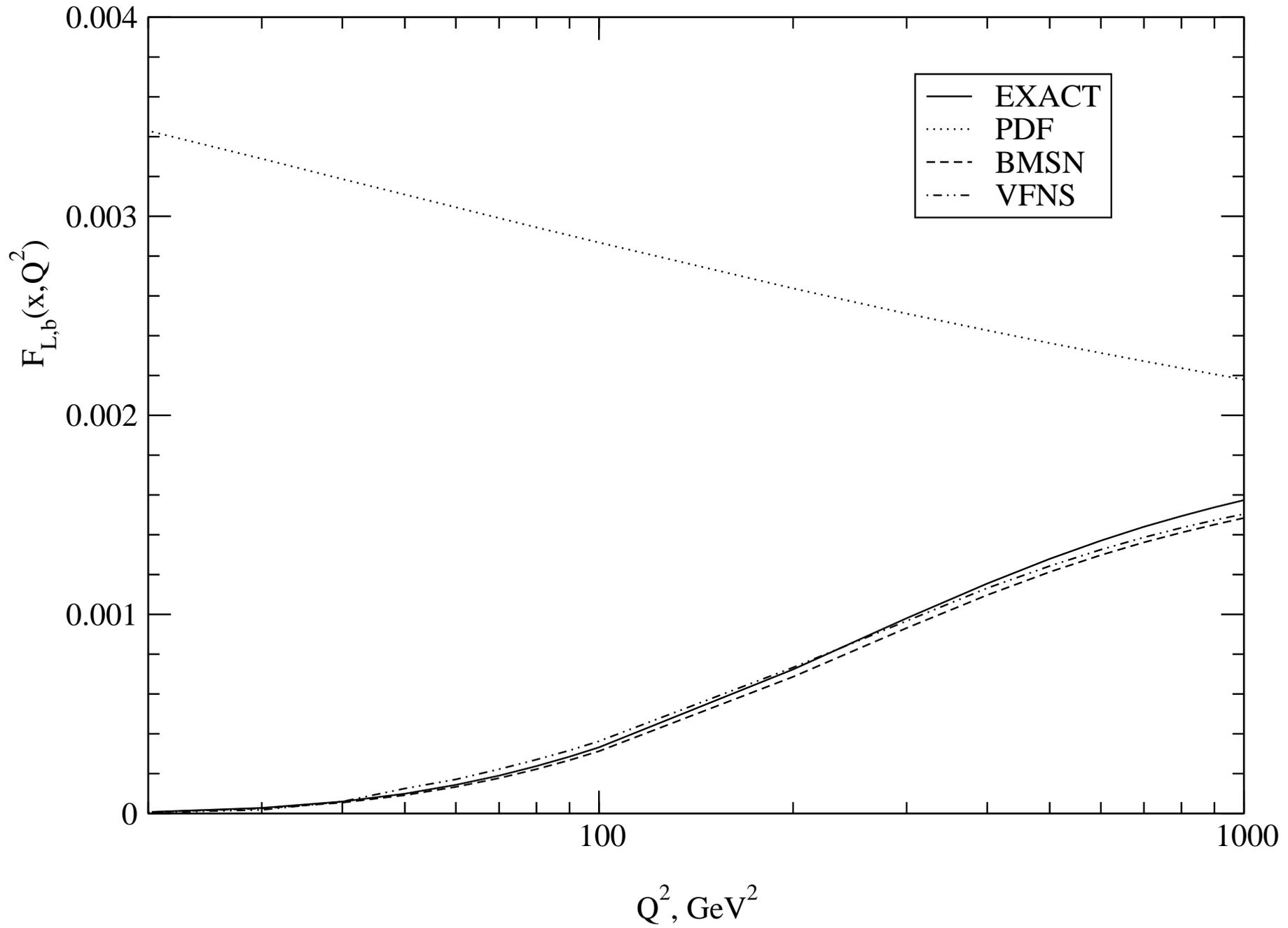
\item[Fig. 3.]
The bottom quark structure functions 
$F_{L,b}^{\rm EXACT}(n_f=4)$ (solid line) 
$F_{L,b}^{\rm CSN}(n_f=5)$, (dot-dashed line)
$F_{L,b}^{\rm BMSN}(n_f=5)$, (dashed line) and
$F_{L,b}^{\rm PDF}(n_f=5)$, (dotted line) 
in NNLO for $x=0.05$
plotted as functions of $Q^2$.
%------------------------------------------------
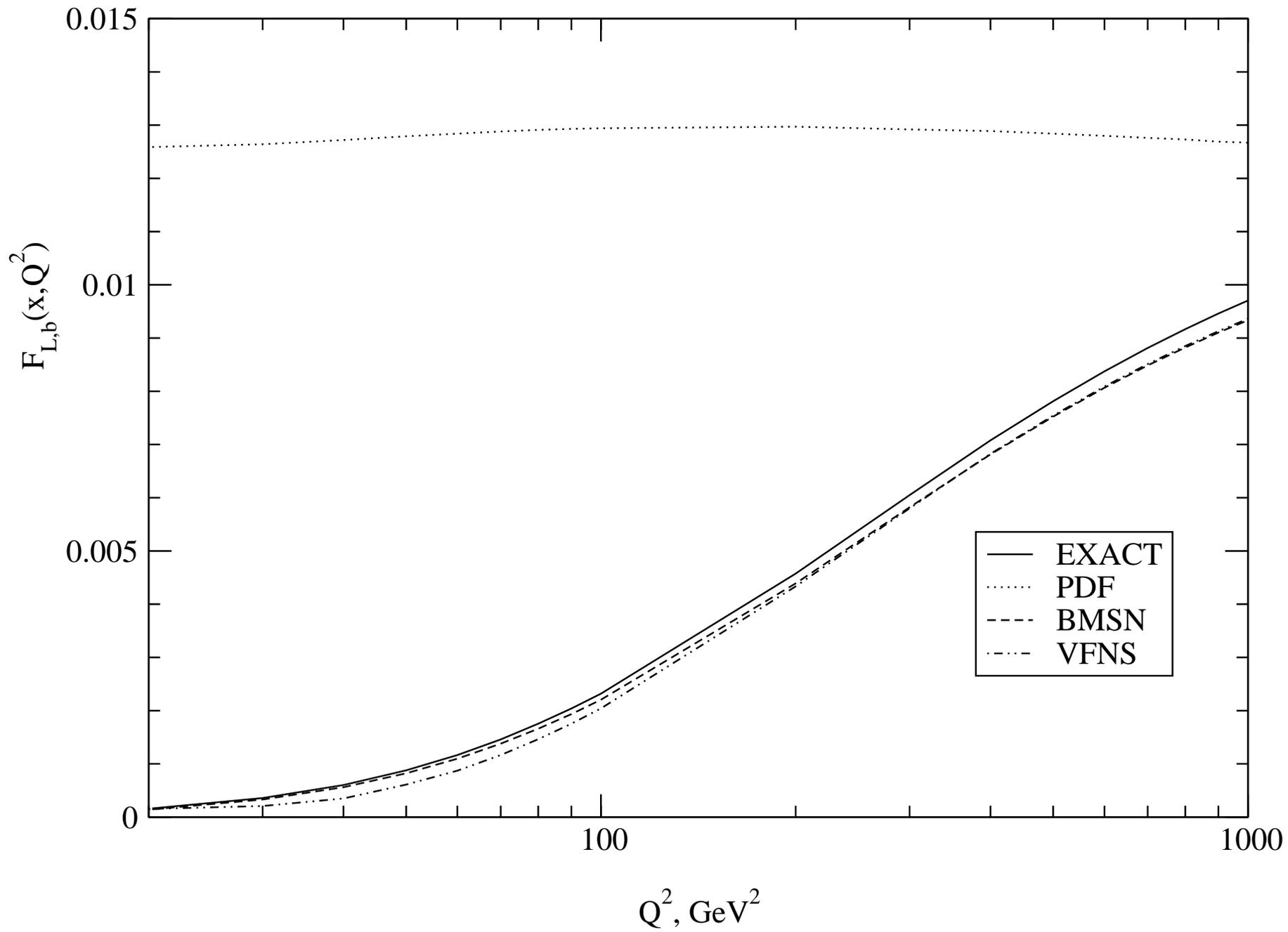
\item[Fig. 4.] Same as in Fig. 3 but now for $x=0.005$.
%---------------------------------------------------
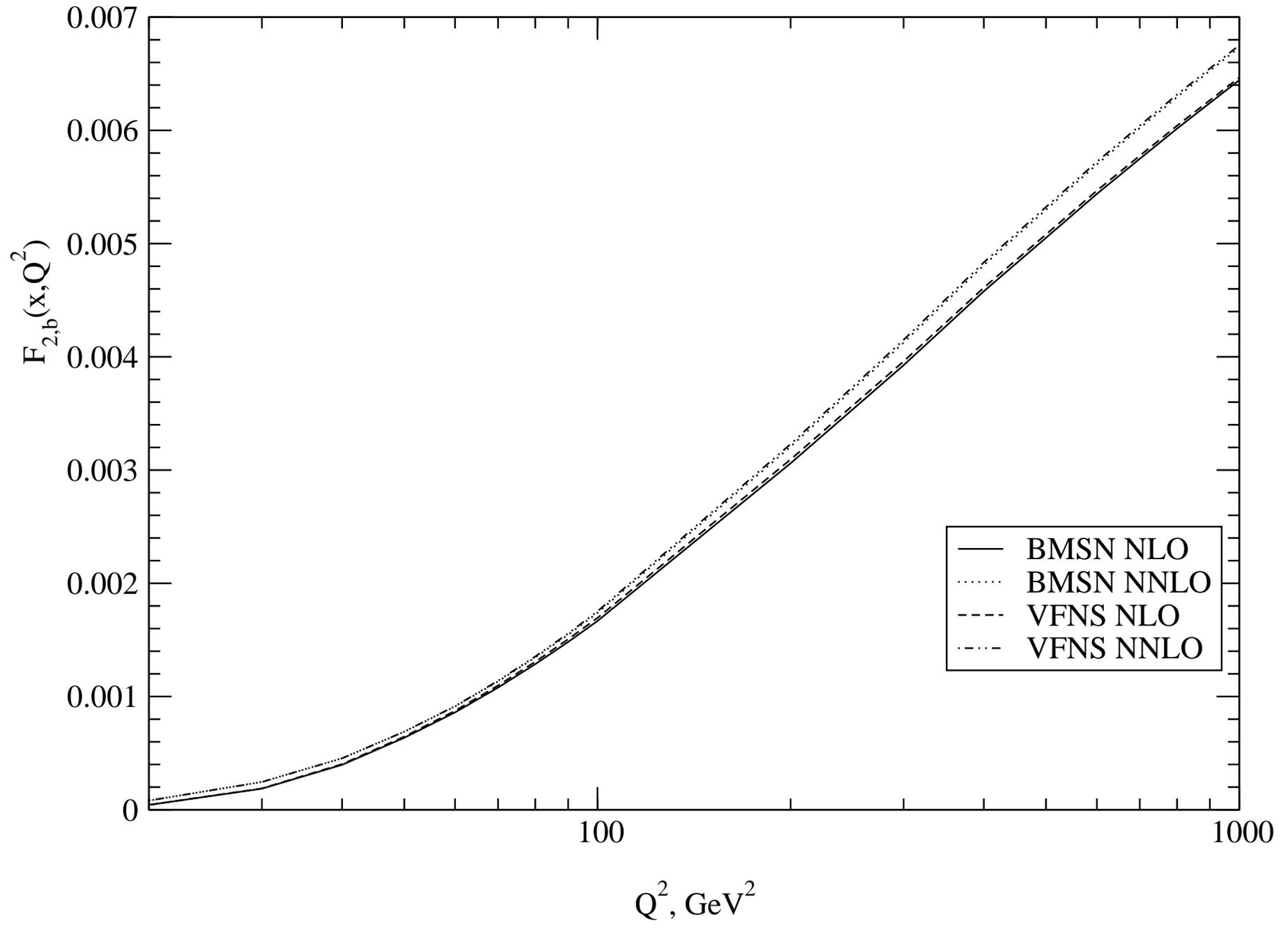
\item[Fig. 5.] 
The bottom quark structure functions 
$F_{2,b}^{\rm BMSN}(n_f=5)$ in NLO (solid line), NNLO 
(dotted line) and 
$F_{2,b}^{\rm CSN}(n_f=5)$ in NLO (dashed line), 
NNLO (dot-dashed line) for $x=0.05$
plotted as functions of $Q^2$.
%------------------------------------------------------
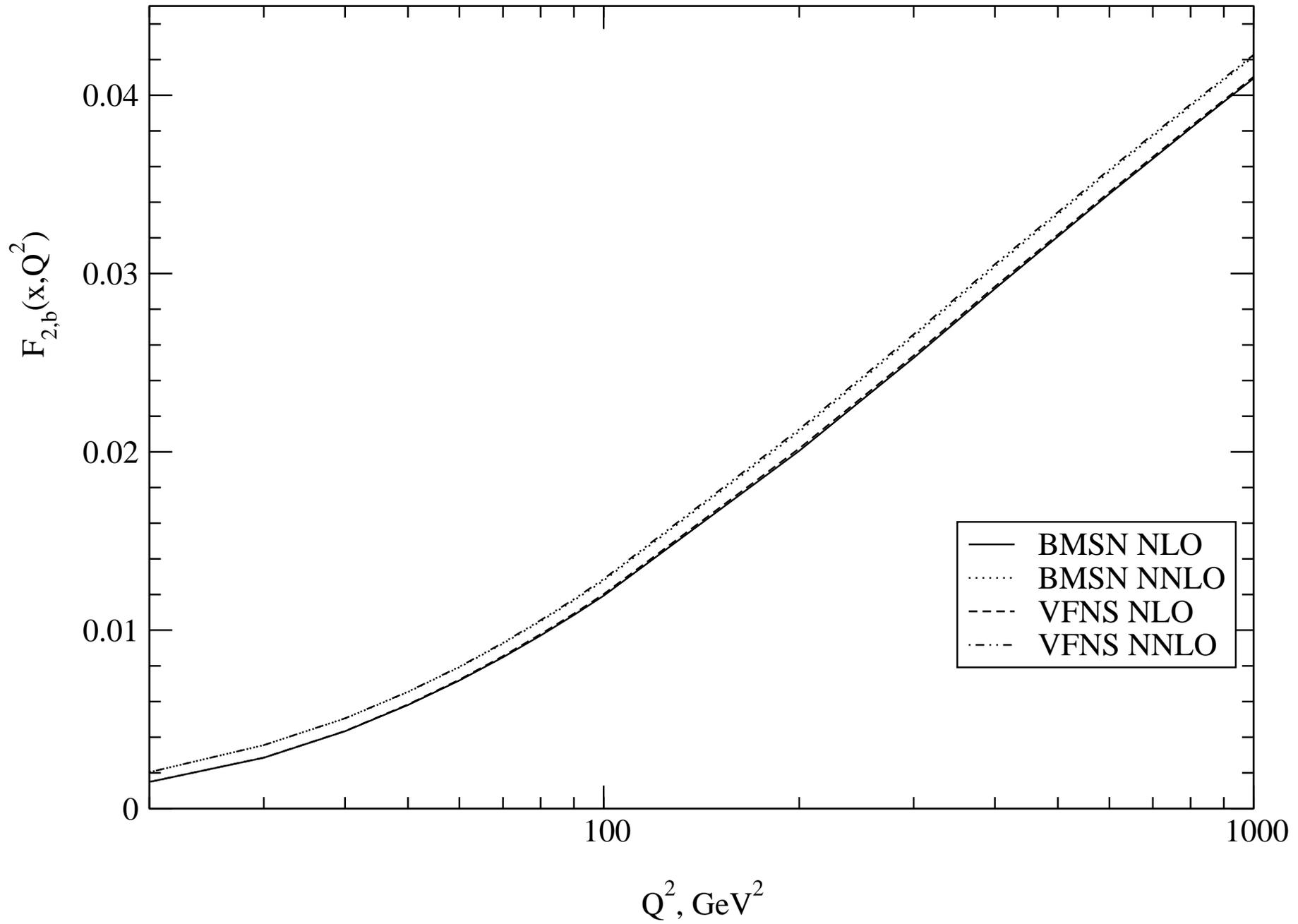
\item[Fig. 6.] Same as in Fig. 5 but now for $x=0.005$.
%-----------------------------------------------------
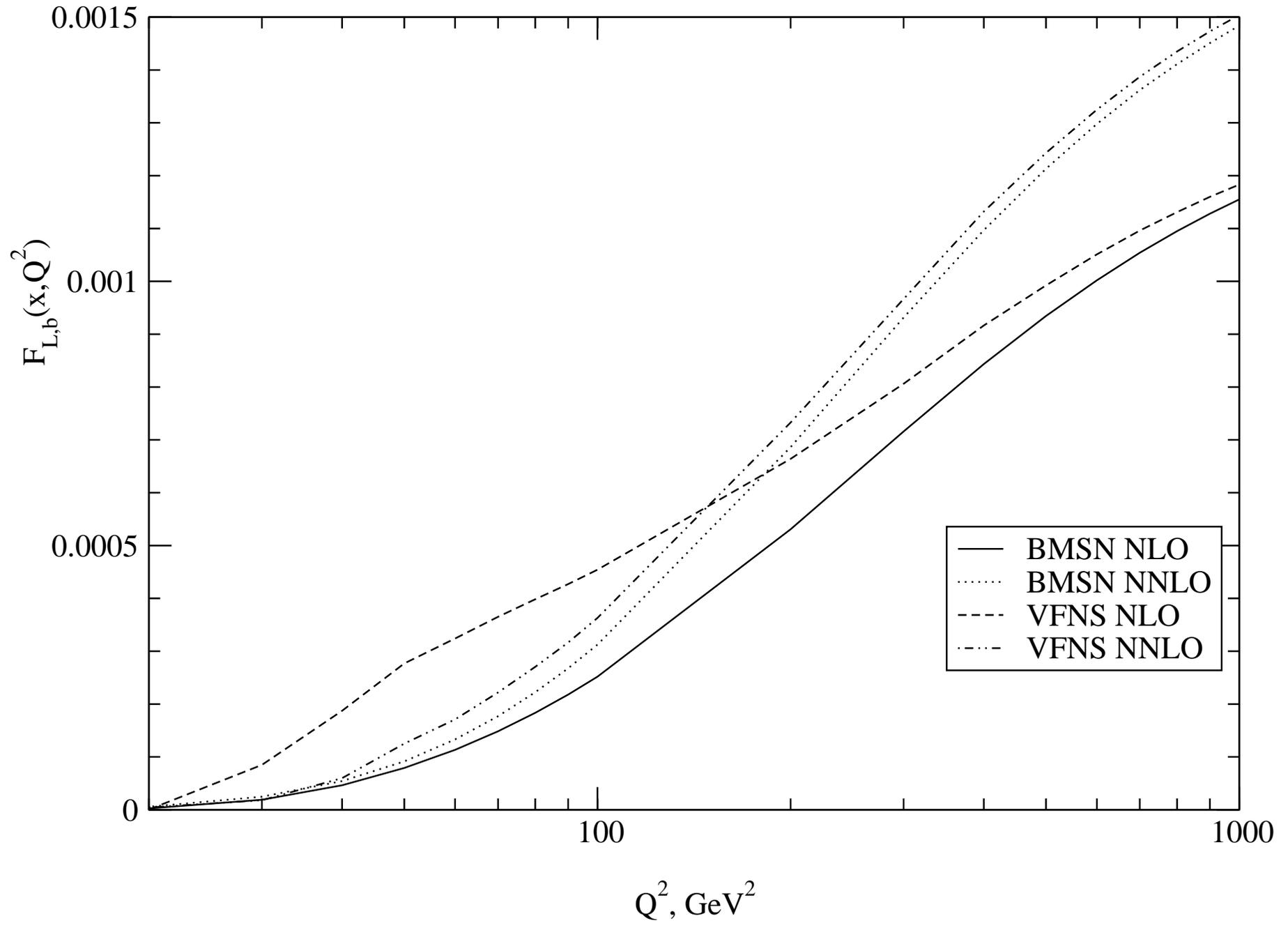
\item[Fig. 7.]
The bottom quark structure functions 
$F_{L,b}^{\rm BMSN}(n_f=5)$ in NLO (solid line), NNLO 
(dotted line) and
$F_{L,b}^{\rm CSN}(n_f=5)$ in NLO (dashed line),
NNLO (dot-dashed line) for $x=0.05$ 
plotted as functions of $Q^2$.
%----------------------------------------------------
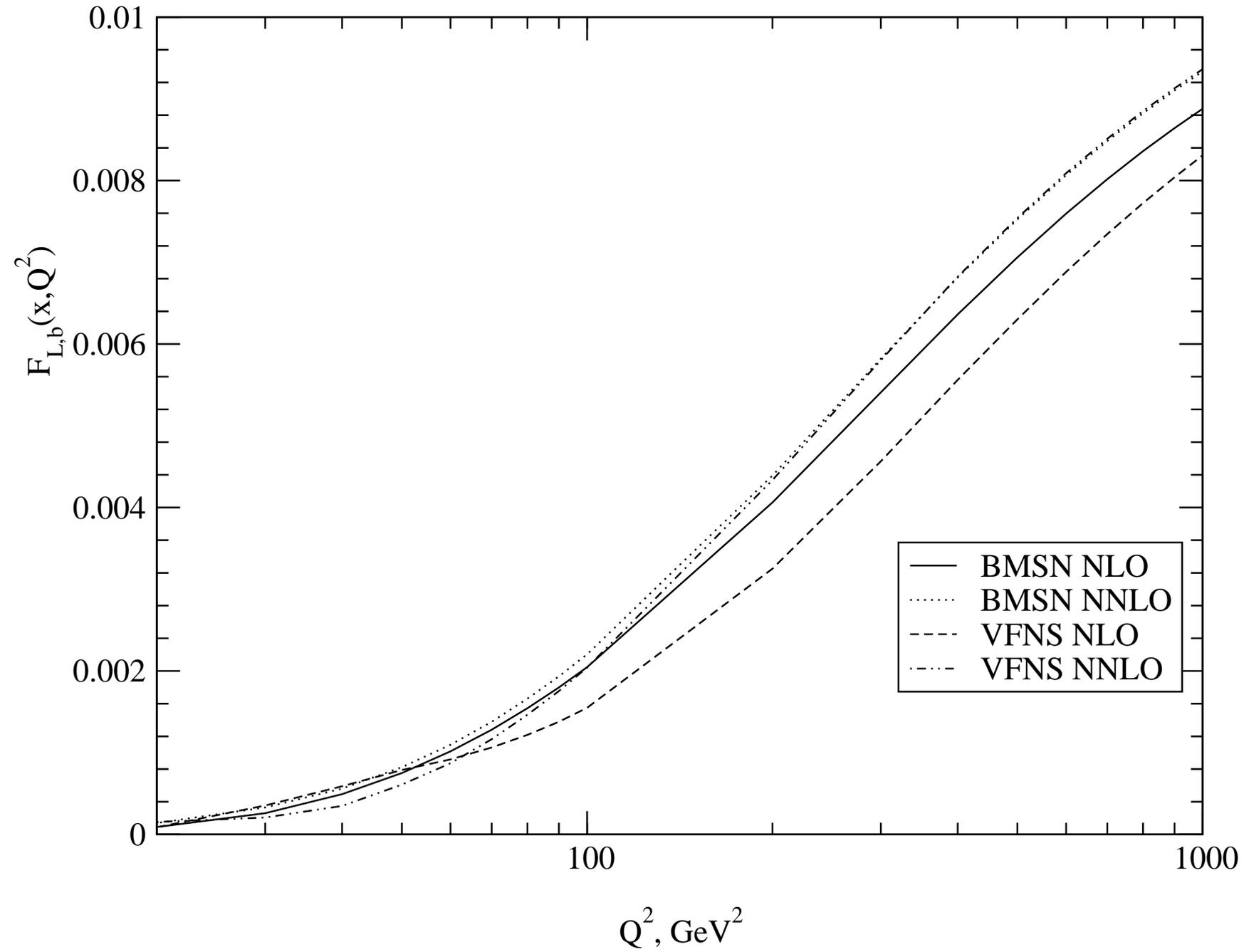
\item[Fig. 8.] Same as in Fig. 7 but now for $x=0.005$.
%-----------------------------------------------------
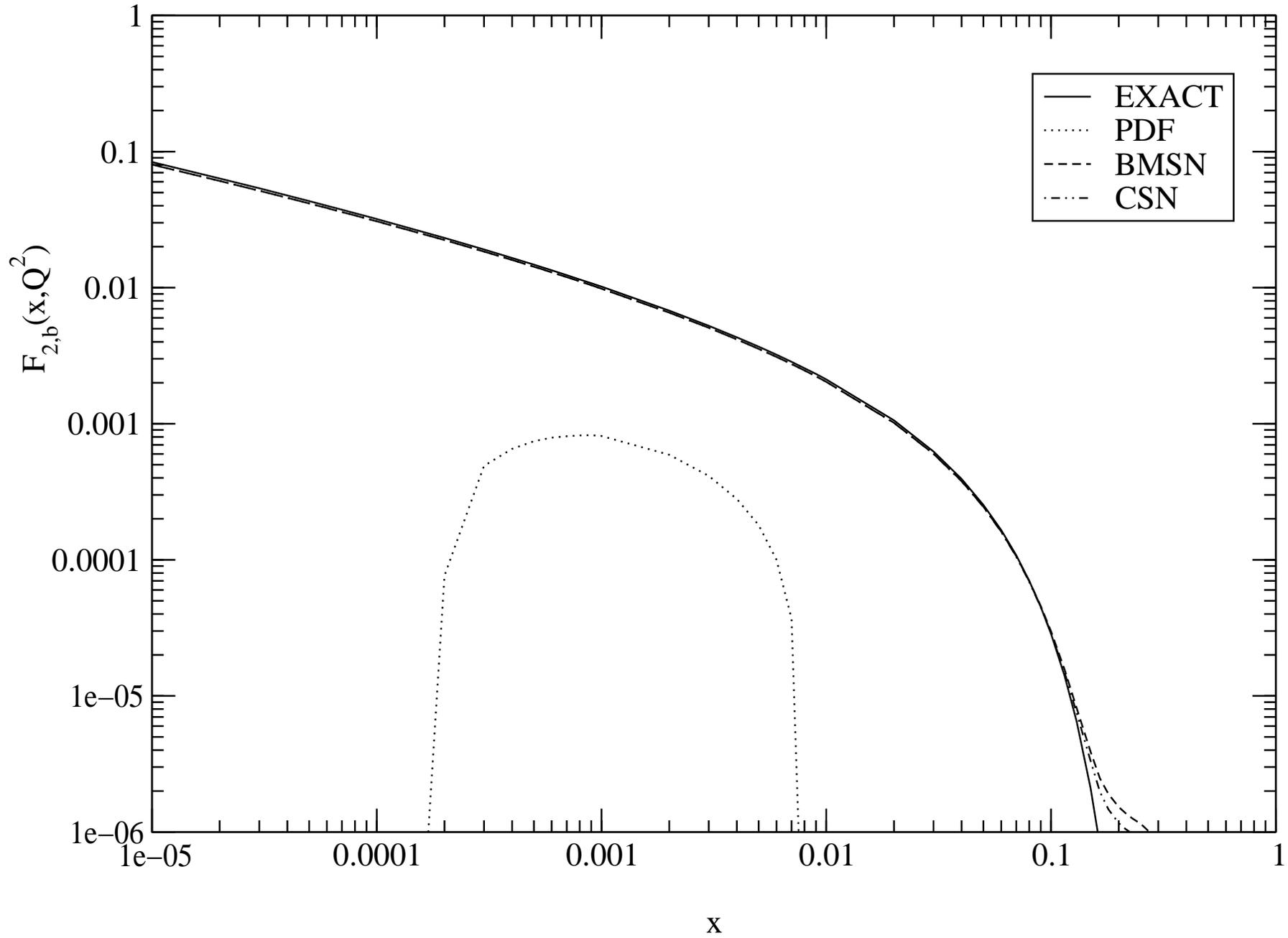
\item[Fig. 9.]
The bottom quark structure functions 
$F_{2,b}^{\rm EXACT}(n_f=4)$ (solid line) 
$F_{2,b}^{\rm CSN}(n_f=5)$, (dot-dashed line)  
$F_{2,b}^{\rm BMSN}(n_f=5)$, (dashed line) and
$F_{2,b}^{\rm PDF}(n_f=5)$, (dotted line) 
in NNLO for $Q^2 = 30$ $({\rm GeV/c})^2$ 
plotted as functions of $x$.
%----------------------------------------------------
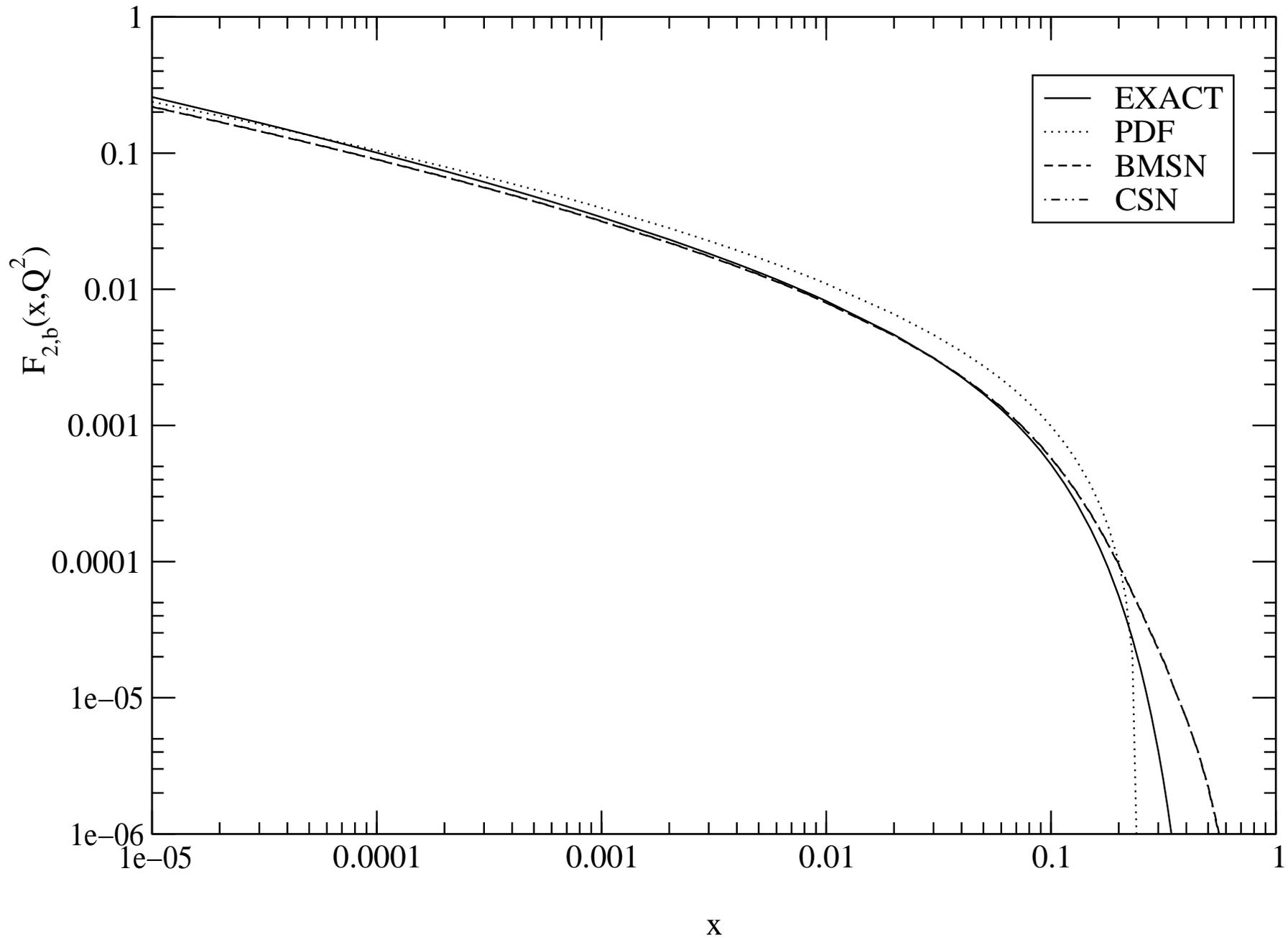
\item[Fig. 10.] Same as in Fig. 9 but now for $Q^2 = 100$
$({\rm GeV/c})^2$.
%---------------------------------------------------------
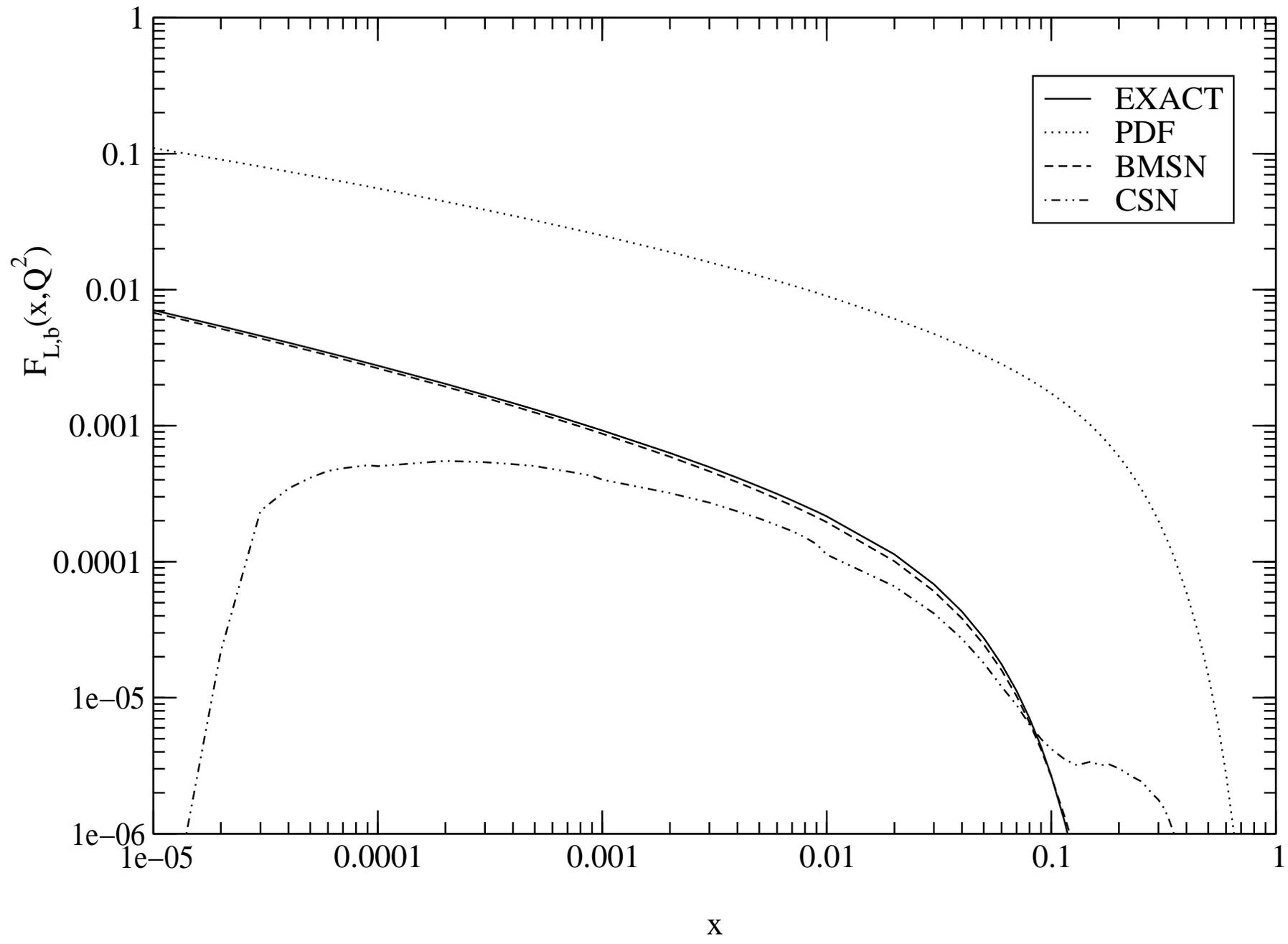
\item[Fig. 11.]
The bottom quark structure functions 
$F_{L,b}^{\rm EXACT}(n_f=4)$ (solid line) 
$F_{L,b}^{\rm CSN}(n_f=5)$, (dot-dashed line)
$F_{L,b}^{\rm BMSN}(n_f=5)$, (dashed line) and
$F_{L,b}^{\rm PDF}(n_f=5)$, (dotted line) 
in NNLO for $Q^2 = 30$ $({\rm GeV/c})^2$ 
plotted as functions of $x$.
%----------------------------------------------------
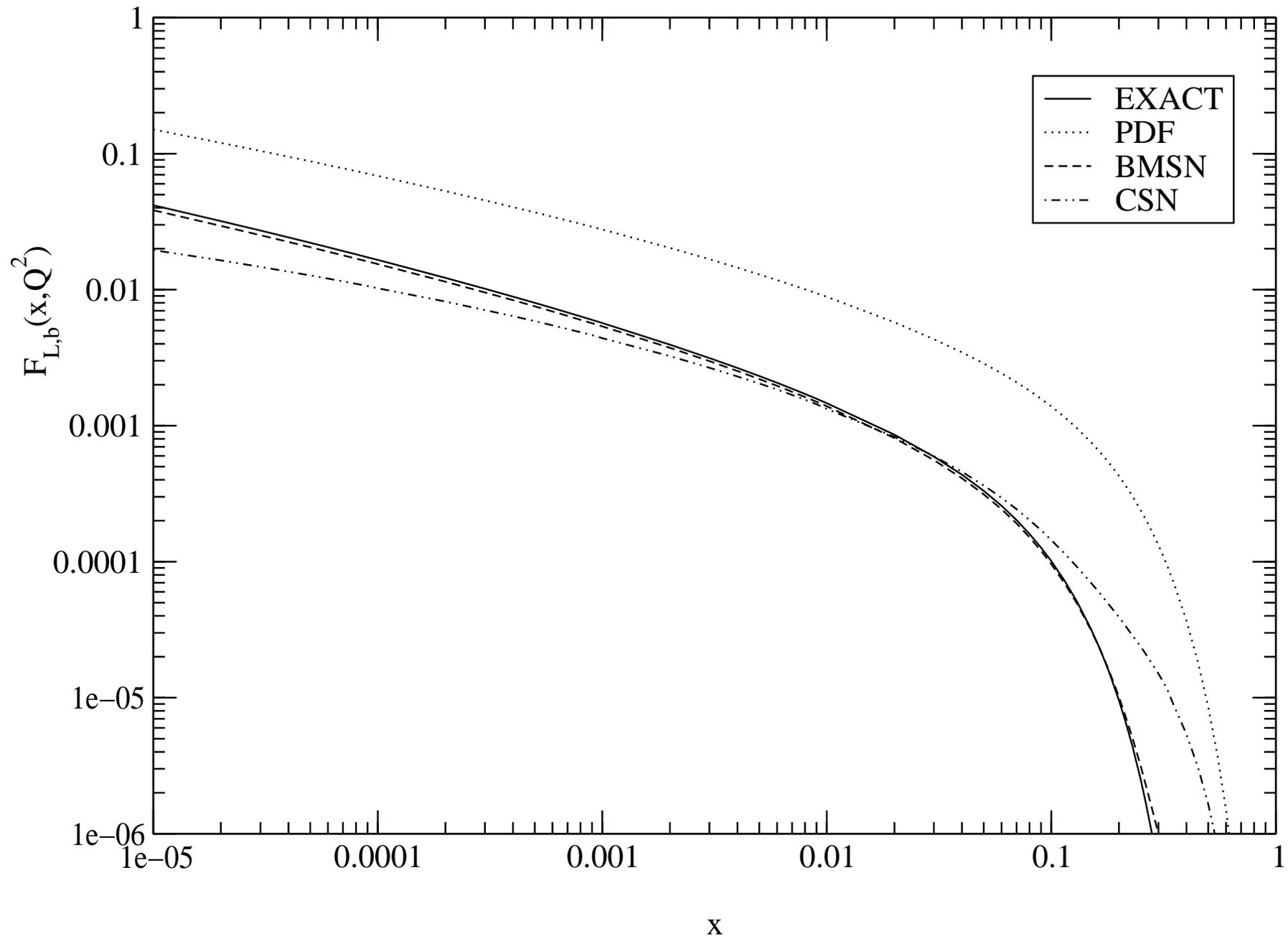
\item[Fig. 12.] Same as in Fig. 11 but now for $Q^2 = 100$
$({\rm GeV/c})^2$.
%---------------------------------------------------------
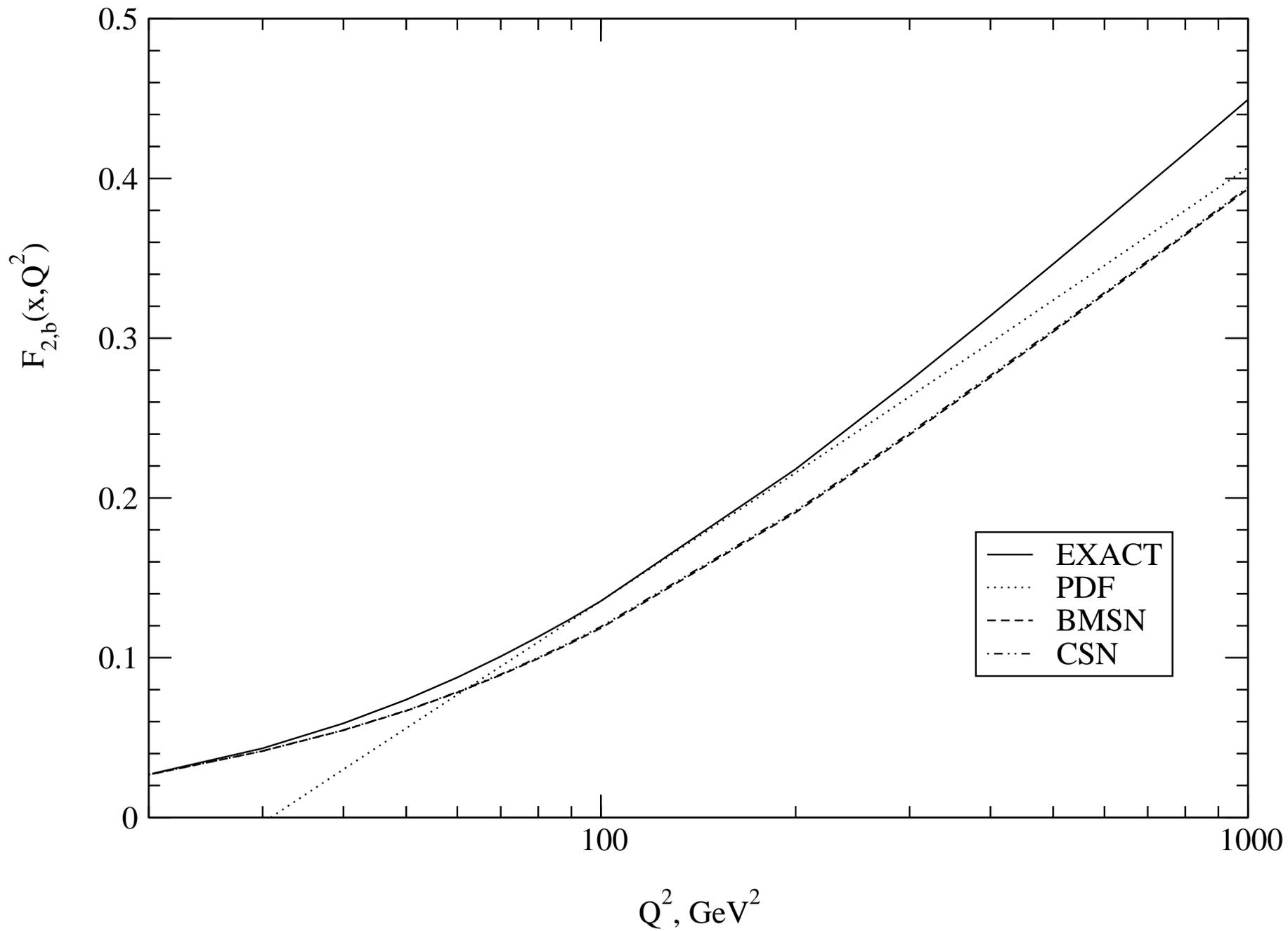
\item[Fig. 13.] Same as in Fig. 1 but now for $x=5\times 10^{-5}$.
%---------------------------------------------------------
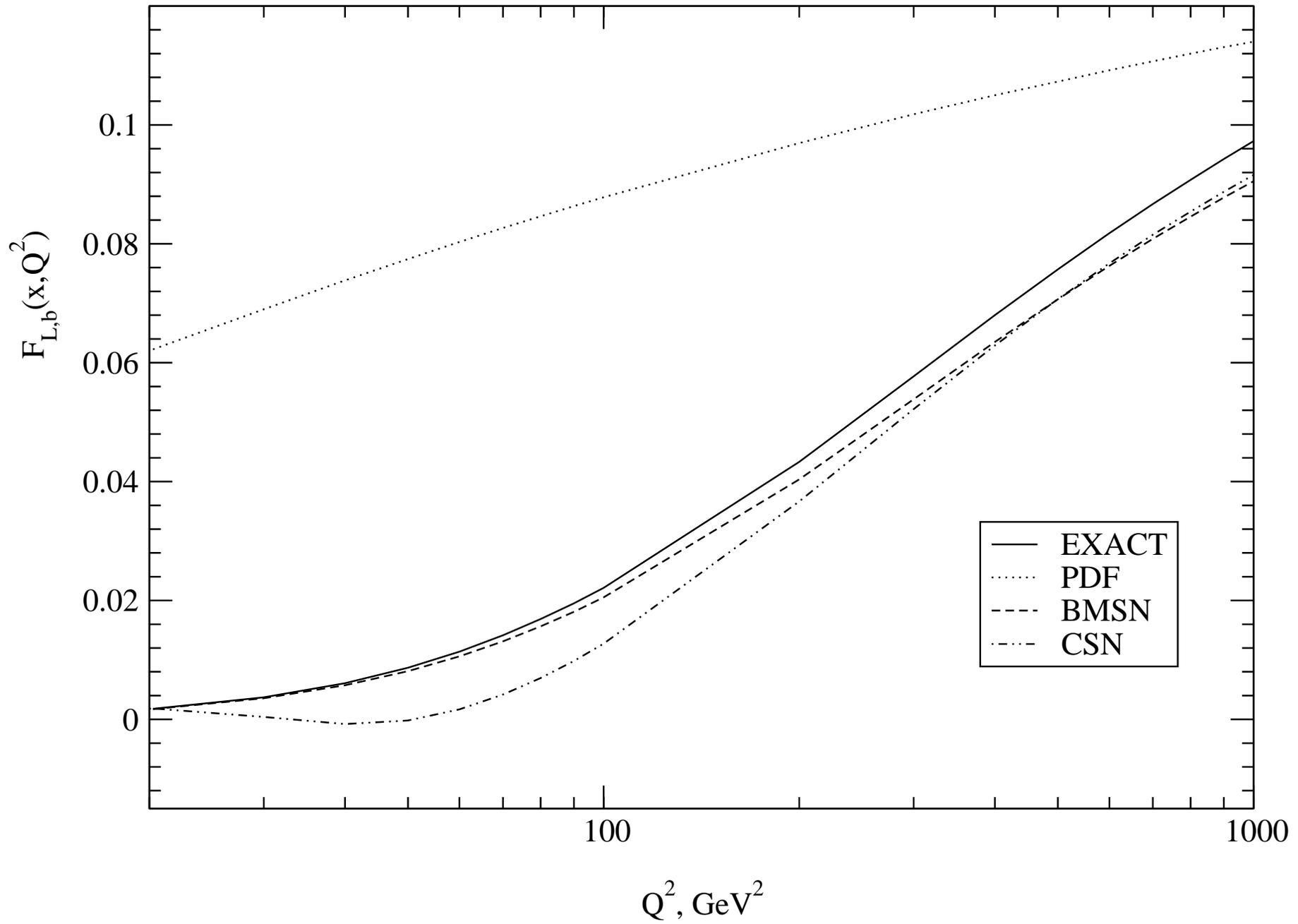
\item[Fig. 14.] Same as in Fig. 3 but now for $x=5\times 10^{-5}$.
%---------------------------------------------------------
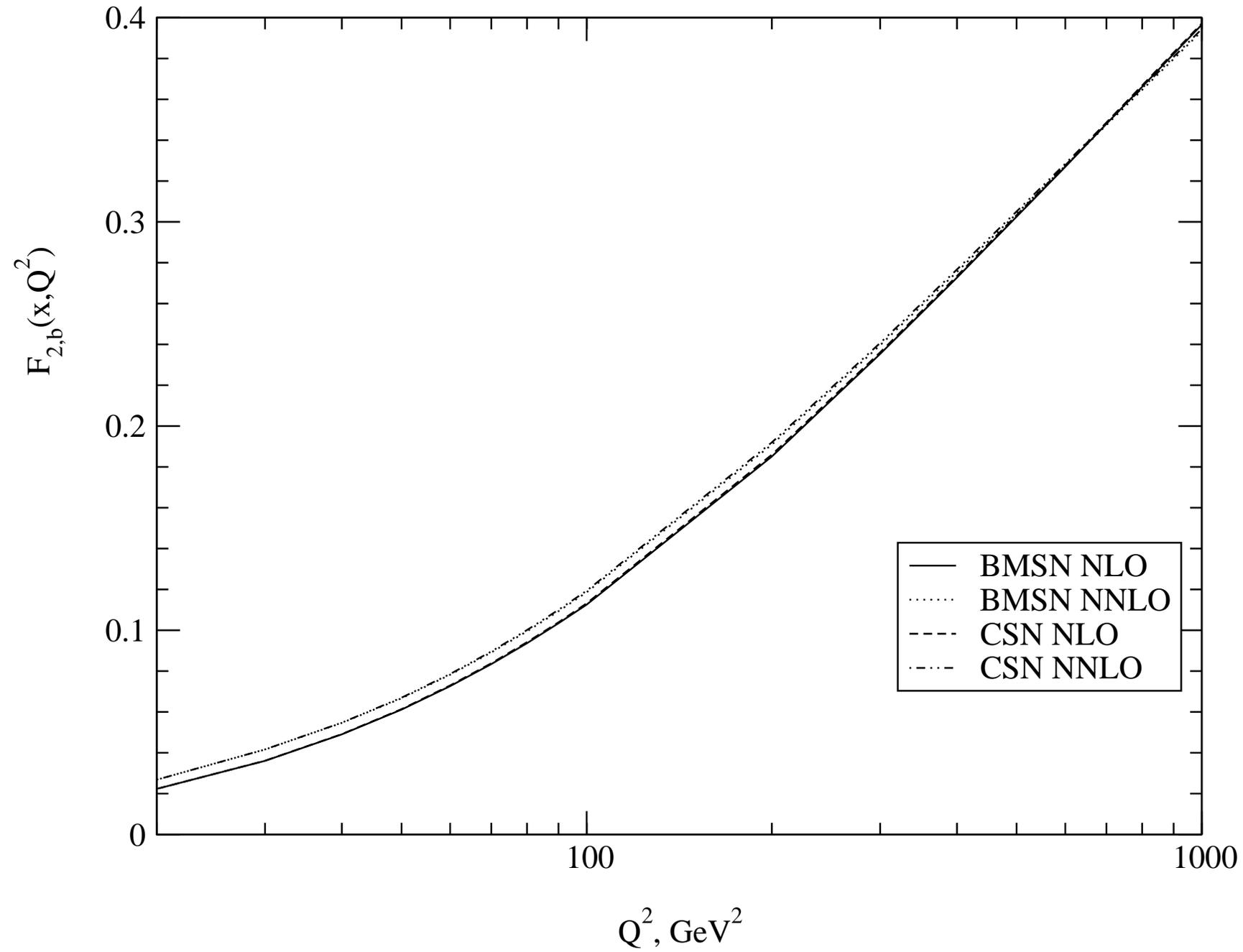
\item[Fig. 15.] Same as in Fig. 5 but now for $x=5\times 10^{-5}$.
%---------------------------------------------------------
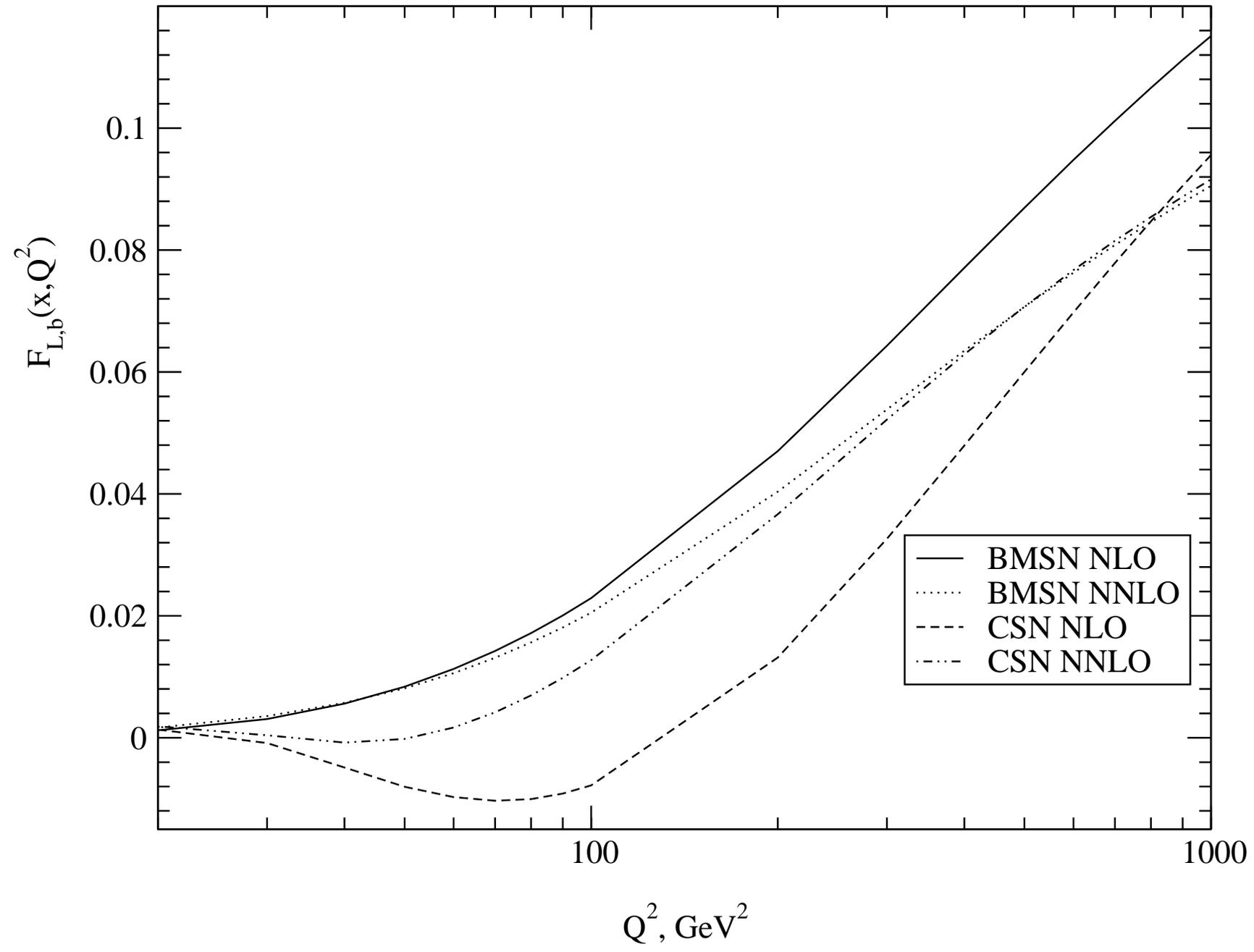
\item[Fig. 16.] Same as in Fig. 7 but now for $x=5\times 10^{-5}$.
%----------------------------------------------------
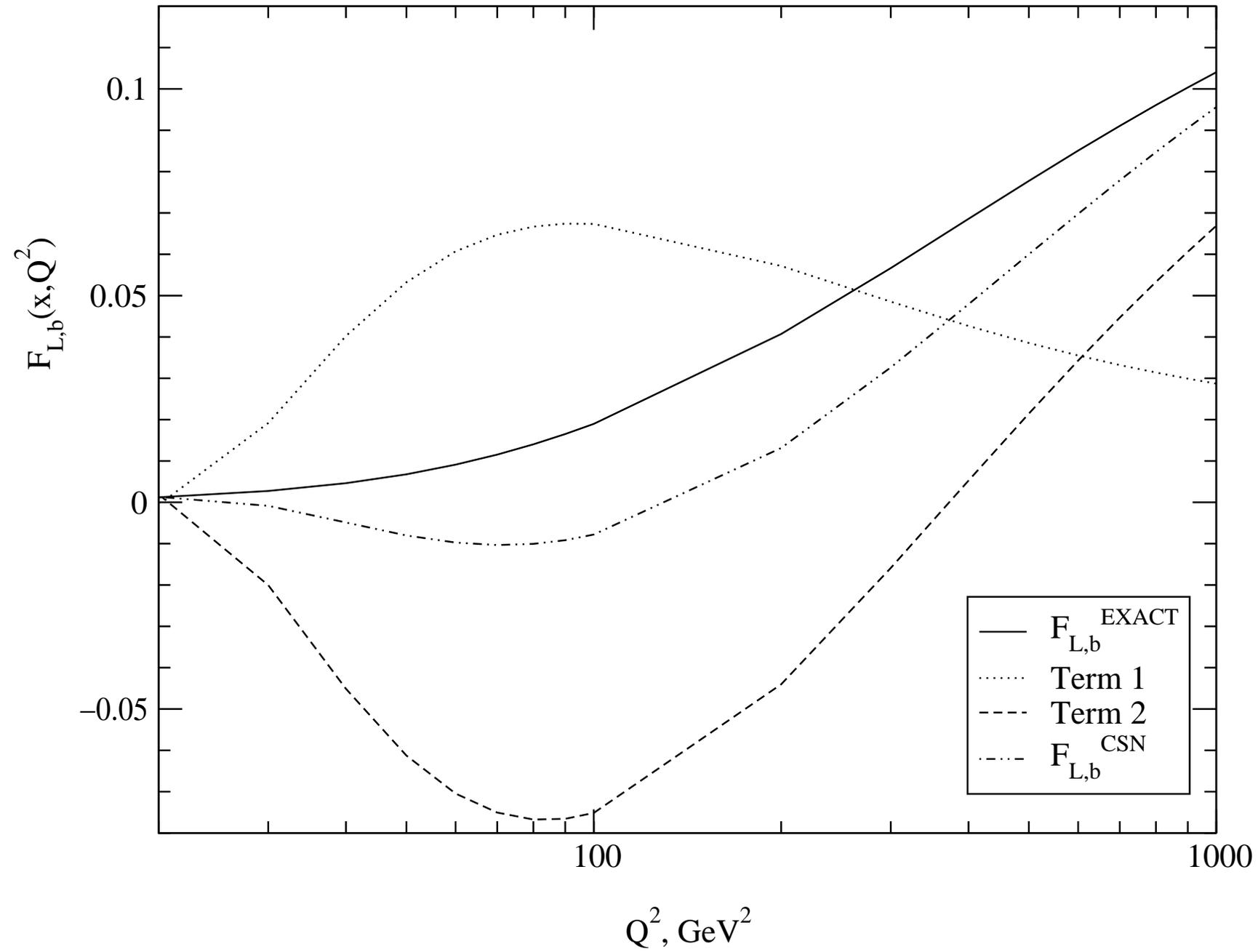
\item[Fig. 17.]
The bottom quark structure function 
$F_{L,b}^{\rm EXACT}(n_f=4)$ (solid line) 
$F_{L,b}^{\rm CSN}(n_f=5)$, (dot-dashed line)  
together with the NLO charm density piece Term1 (dotted line) and
the LO gluon density piece Term2 (dashed line), see text,
for $x=5\times 10^{-5}$ plotted as functions of $Q^2$.
%--------------------------------------------------------------
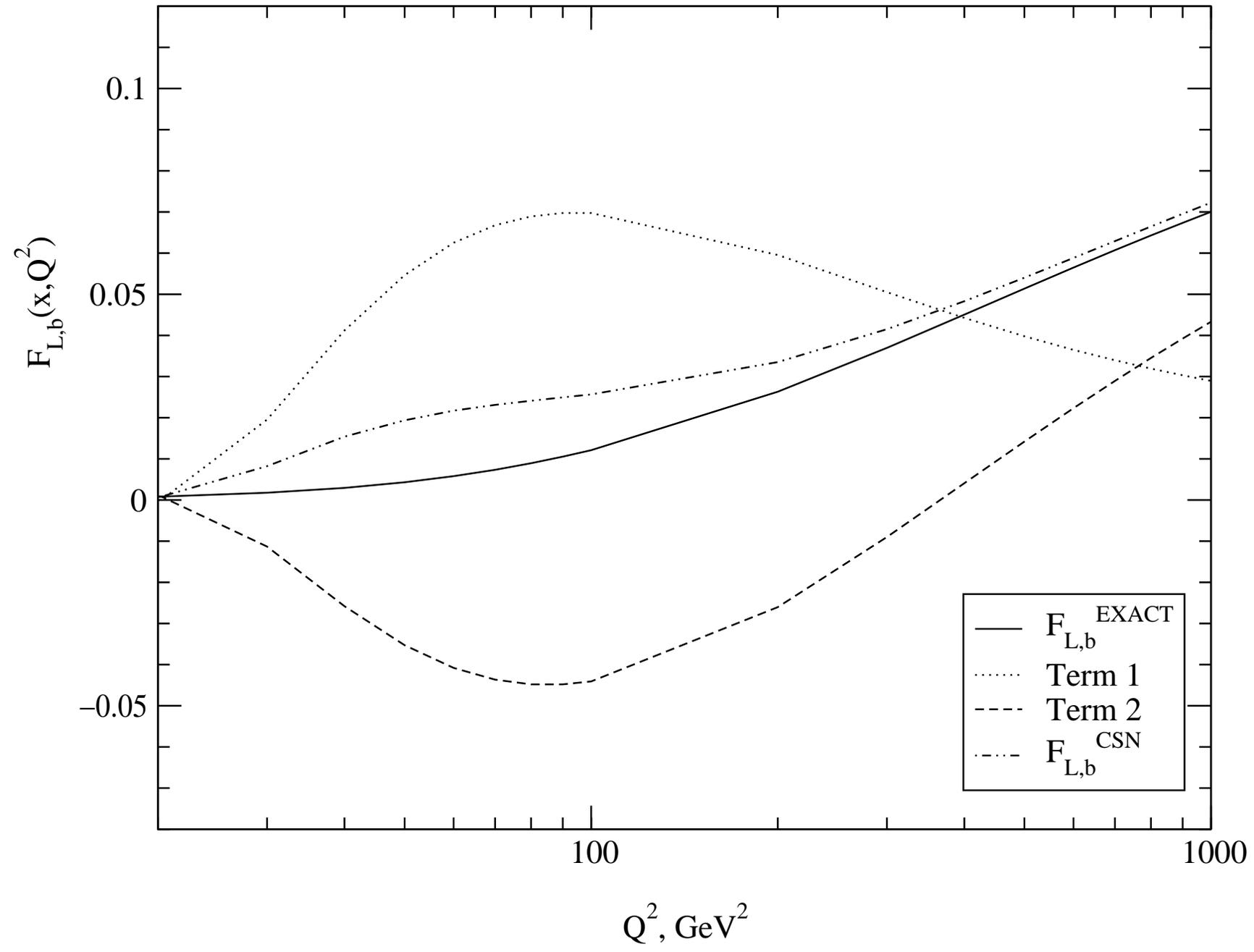
\item[Fig. 18.]
The bottom quark structure function 
$F_{L,b}^{\rm EXACT}(n_f=4)$ (solid line) 
$F_{L,b}^{\rm CSN}(n_f=5)$, (dot-dashed line)  
together with the NLO charm density piece Term1 (dotted line) and
the NLO gluon density piece Term2 (dashed line), see text,
for $x=5\times 10^{-5}$ plotted as functions of $Q^2$.
%-------------------------------------------------------------
\end{description}
%\end{document}

%------------------------------
\end{document}